\documentclass[twocolumn,showpacs,preprintnumbers,nofootinbib,amsmath,amssymb]{revtex4}
\usepackage{amsmath}
\usepackage{euscript,amssymb}
\usepackage{epsfig}
\usepackage{graphicx}
\usepackage{amssymb}

\usepackage{amsfonts,latexsym}
\usepackage{euscript,amssymb}
\usepackage{epsfig}

\def\ZZ{{\mathbb Z}}

\newcommand{\be}[1]{\begin{equation}\label{#1}}
\newcommand{\ee}{\end{equation}}
\newcommand{\ba}[1]{\begin{eqnarray}\label{#1}}
\newcommand{\ea}{\end{eqnarray}}
\newcommand{\rf}[1]{(\ref{#1})}
\newcommand{\nn}{\nonumber}

\newcommand{\const}{\mbox{\rm const}\,}
\newcommand{\sign}{ \mbox{\rm sign}\,}
\newcommand{\opensquare}{\mbox{$\rlap{$\sqcap$}\sqcup$}}

\renewcommand{\theequation}{\arabic{section}.\arabic{equation}}

\begin{document}

\title{Bouncing inflation in nonlinear $R^2+R^4$ gravitational model}

\author{Tamerlan Saidov}\email{tamerlan-saidov@yandex.ru}  \author{Alexander Zhuk}\email{ai_zhuk2@rambler.ru}
\affiliation{Astronomical Observatory and Department of
Theoretical Physics, Odessa National University, 2 Dvoryanskaya
Street, Odessa 65082, Ukraine}


\begin{abstract}We study a gravitational model
with curvature-squared $R^2$ and curvature-quartic $R^4$ nonlinearities. The effective scalar degree of freedom $\phi$ (scalaron) has a multi-valued potential $U(\phi)$ consisting of a number of branches. These branches are fitted with each other in the branching and monotonic points.
In the case of four-dimensional space-time, we show that the monotonic points are penetrable for scalaron while in the vicinity of the branching points scalaron has the bouncing behavior and cannot cross these points. Moreover, there are branching points
where scalaron bounces  an infinite number of times with decreasing amplitude and the Universe asymptotically approaches the de Sitter stage. Such accelerating behavior we call bouncing inflation. For this accelerating expansion there is no need for original potential $U(\phi)$ to have a minimum or to check the slow-roll conditions. A necessary condition for such inflation is the existence of the branching points. This is a new type of inflation. We show that bouncing inflation takes place both in the Einstein and Brans-Dicke frames.
\end{abstract}


\maketitle


\section{\label{sec:1}Introduction}

\setcounter{equation}{0}

Starting from the pioneering paper \cite{Star1}, the nonlinear (with respect to the
scalar curvature $R$) theories of gravity $f(R)$
have attracted the great deal of interest because these models can provide a natural mechanism
of the early inflation.
Nonlinear models may arise either due to quantum fluctuations of matter fields including gravity \cite{BirrDav}, or as a result of compactification
of extra spatial dimensions \cite{NOcompact}. Compared, e.g., to others higher-order gravity theories,
$f(R)$ theories are free of ghosts and of Ostrogradski instabilities \cite{Woodard}.
Recently, it was realized that these models can also explain the late time acceleration
of the Universe. This fact resulted in a new wave of papers devoted to this topic (see e.g., recent reviews \cite{reviews,CLF}).

The most simple, and, consequently, the most studied models are
polynomials of $R$:  $f(R)=\sum_{n=0}^k C_n R^n \,$ $(k >1)$, e.g., quadratic $R+R^2$ and quartic $R+R^4$ ones. Active investigation of these models, which started in 80-th years of the last century \cite{80-th,Maeda}, continues up to now \cite{Ketov}. Obviously, the correction terms (to the Einstein action) with $n>1$ give the main contribution in the case of large $R$, e.g., in the early stages of the Universe evolution. As it was shown first in \cite{Star1} for the quadratic model, such modification of gravity results in early inflation. From the other hand,
function $f(R)$ may also contain negative degrees of $R$. For example, the simplest model is $R+R^{-1}$. In this case the correction term plays the main role for small $R$, e.g., at the late
stage of the Universe evolution (see e.g. \cite{GZBR,SZPRD2007} and numerous references therein). Such modification of gravity may result in the late-time acceleration of our Universe \cite{Carrolletal}.
Nonlinear models with polynomial as well as $R^{-1}$-type correction terms have also been generalized to the multidimensional case (see e.g.,  \cite{GZBR,SZPRD2007,Ellis,GMZ1,GMZ2,SZGC2006,Bronnikov,SZPRD2009}).

It is well known that nonlinear models are equivalent to linear-curvature
models with additional scalar field $\phi$ (dubbed scalaron in \cite{Star1}). This scalar field corresponds
to additional degree of freedom of nonlinear models. The dynamics of this field (as well as a possibility of inflation of the Universe) is defined by potential $U (R(\phi ),\phi)$ (see Eq. \rf{1.8} below\footnote{Starting from Sec. II, we denote the scalar curvature of the original nonlinear model by $\bar R$.}) where $R=R(\phi )$ is a solution of Eq. \rf{1.7}:  $\exp{(A\phi)} = df/dR$. Usually, models or a particular cases of these models are considered where this equation has only one solution. In this case, potential $U$ is a one-valued function of $\phi $. However, in the most general case this equation has more than one solution and potential becomes a multi-valued function with a number of branching points (see e.g., \cite{Frolov}). Investigation of the dynamical behavior of scalar field and Universe in such models (especially in the vicinity of the branching points) is not a trivial problem and may result in new important effects. Therefore, it is of interest to consider the models with multi-valued potentials.

In the present paper, we study an example of such models. Here, $f(R)$ has both quadratic $R^2$ and quartic $R^4$ contributions. In this case Eq. \rf{1.7} is a cubic equation with respect to $R$ and may have, in general, three real solutions/branches $R_{i} (\phi)\, (i=1,2,3)$. We have investigated this model in our paper \cite{SZGC2006}. However, in this paper we considered a special case of one real solution in $D=8$ space-time. Now, we study the most interesting case of three real solutions. These solutions are fitted with each other in the branching points. There are also another type of matching points where one-valued solutions are fitted with  the three-valued solutions. In the vicinity of these points potential $U(\phi )$ is a monotonic function. Thus, these latter points dubbed monotonic ones. The main aim of the paper consists in the investigation of  dynamical behavior of the system in four dimensional space-time in the vicinity of the branching and monotonic points. We show that dynamics is quite different for branching and monotonic points. The monotonic points are penetrable for scalaron while in the branching points scalaron has bouncing behavior. There are branching points where scalaron bounces  an infinite number of times with decreasing amplitude and the Universe asymptotically approaches the de Sitter stage. Such accelerating behavior we call {\em bouncing inflation}. We should note that for this type of inflation there is no need for original potential $U(\phi)$ to have a minimum or to check the slow-roll conditions.
A necessary condition for such inflation is the existence of the branching points.
This is a new type of inflation. We show that this inflation takes place both in the Einstein and Brans-Dicke frames. This is the main result of our paper. We think that scalaron field and the Universe have the similar behavior in the vicinity of the branching points for others polynomial and $R^{-1}$-type models resulting in both early inflation and late-time acceleration.
Of course, it is necessary to conduct additional studies of these models, to confirm or refute this assertion.

The paper is structured as follows. In Sec. II we study briefly the equivalence between an arbitrary nonlinear $f(\bar R)$ theory and theory linear in another scalar
curvature $R$ but which contains a scalaron field $\phi $. In Sec. III we consider a particular example of nonlinear model with curvature-quadratic and curvature-quartic correction terms and obtain solutions/branches $\bar R_{i}(\phi )$. The fitting procedure for these branches is proposed in Sec. IV. The dynamics of the scalaron and the Universe is investigated in Sec. V. Here, we parameterize the scalaron potential in such a way that it becomes a one-valued function. It gives a possibility to study the dynamical behavior of the system in the vicinity of the branching and monotonic points. We show that in the vicinity of the branching point the scalaron field bounces  an infinite number of times with decreasing amplitude and the
Universe acquires the accelerating expansion approaching asymptotically  to the de Sitter stage. Such accelerating expansion we call bouncing inflation. A brief discussion of the obtained results is presented in the
concluding Sec. VI. In Appendix A, we show that bouncing inflation in the vicinity of the branching point takes place also in the Brans-Dicke frame.


\section{\label{setup}General setup}
\setcounter{equation}{0}

It is well known that nonlinear theories
\be{1.1} S = \frac {1}{2\kappa^2_D}\int_M d^Dx \sqrt{|\bar g|}
f(\bar R)\; , \ee
where $f(\bar R)$ is an arbitrary smooth function of a
scalar curvature $\bar R = R[\bar g]$ constructed from the
$D-$dimensional metric $\bar g_{ab}\; (a,b = 1,\ldots,D)$
are equivalent
to  theories which are linear in another scalar
curvature $R$ but which contains an additional self-interacting
scalar field. According to standard techniques
\cite{80-th,Maeda},
the corresponding $R-$linear theory has the
action functional:
\be{1.6} S = \frac{1}{2\kappa^2_D} \int_M d^D x \sqrt{|g|} \left[
R[g] - g^{ab} \phi_{,a} \phi_{,b} - 2 U(\phi )\right]\; , \ee
where
\be{1.7} f'(\bar R) = \frac {df}{d \bar R} := e^{A \phi} > 0\; ,\quad A :=
\sqrt{\frac{D-2}{D-1}}\;  ,\ee
and where the self-interaction potential $U(\phi )$ of the scalar
field $\phi$ is given by
\ba{1.8-1} U(\phi ) &=&
 \frac12 \left(f'\right)^{-D/(D-2)} \left[\; \bar R f' - f\right]\;
 ,\label{1.8a}\\
&=& \frac12 e^{- B \phi} \left[\; \bar R (\phi )e^{A \phi } -
f\left( \bar R (\phi )\right) \right]\; , \\ \quad B &:=& \frac
{D}{\sqrt{(D-2)(D-1)}}\, .\label{1.8}
\ea
The metrics $g_{ab}$, $\bar g_{ab}$ and the scalar curvatures $R$,
$\bar R$ of the two theories \rf{1.1} and \rf{1.6} are conformally
connected by the relations\footnote{The metrics $g_{ab}$ and $\bar g_{ab}$ represent the Einstein and Brans-Dicke frames, respectively.\label{frames}}
\be{1.9} g_{ab} = \Omega^2 \bar g_{ab} = \left[ f'(\bar
R)\right]^{2/(D-2)}\bar g_{ab}\;  \ee
and
\ba{1.10} R &=& (f')^{2/(2-D)}\left\{ \bar R +\frac{D-1}{D-2}
(f')^{-2} \bar g^{ab}\partial_a f'\partial_b f' \right.\nn\\
 &-& \left. 2\frac{D-1}{D-2}(f')^{-1} \bar {\opensquare} f'\right\}\;\ea
via the scalar field $\phi=\ln[f'(\bar R)]/A$. This scalar field
$\phi$, known as scalaron \cite{Star1},  carries an additional degree of freedom of original nonlinear model.

According to our definition
\rf{1.7}, we consider the positive branch
$f'(\bar{R})>0$.
Although the negative $f'<0$ branch can be considered as well (see
e.g. Refs. \cite{Maeda,GZBR,GMZ2}). However, negative values of $f'(\bar R)$
result in negative effective gravitational "constant" $G_{eff}=\kappa^2_D/f'$. Thus $f'$
should be positive for the graviton to carry positive kinetic energy (see e.g., \cite{CLF}).

From action \rf{1.6} we obtain  the equation of motion of scalaron field $\phi$:
\be{1.11a}
\opensquare \phi -\frac{\partial U}{\partial \phi} =0\, .
\ee
If scalaron potential $U(\phi )$ has a minimum in a point $\phi_0$:
\ba{1.12a}
&&\left.\frac{d U}{d\phi}\right|_{\phi_{0}} =
\left.\frac{A}{2(D-2)}\left(f'\right)^{-D/(D-2)}h\right|_{\phi_{0}}
= 0, \nn\\ &&h:=Df-2\bar R f', \quad \Longrightarrow \quad \
h(\phi_0) = 0
\ea
then we can define the mass squared of the scalaron \cite{GZBR}:
\ba{1.13a}
m_{\phi}^2 &=&\left.\frac{d^2 U}{d\phi^2}\right|_{\phi_{0}} =
\left.\frac{1}{2(D-1)f''}\frac{(D-2)f'-2\bar R f''}{(f')^{2/(D-2)}}\right|_{\phi_{0}}\nn\\
&=& \left.\frac{(D-2)f'-Dff''/f'}{2(D-1)f''(f')^{2/(D-2)}}\right|_{\phi_{0}}
>0.
\ea
Similar expression for the scalaron mass squared is also given e.g., in \cite{CLF}. The only difference consists in additional
conformal prefactor $1/(f')^{2/(D-2)}$ originated from the conformal metric transformation\footnote{Conformal transformation for mass squared of scalar fields in
models with conformally related metrics is discussed in \cite{Rio}.\label{conf mass}} \rf{1.9}. Up to this prefactor, the positiveness condition \rf{1.13a} of the mass squared coincides
with the stability condition of de Sitter space in $f(\bar R)$ gravity with respect to inhomogeneous and homogeneous perturbations \cite{CLF,86}.
Additionally, to avoid the Dolgov-Kawasaki instability \cite{DK} (instability with respect to local perturbations), it is also required that $f''(\bar R)\geq 0$ (see also \cite{Star2}).

For further research is useful to introduce a new variable
\be{1.14a}
X :=e^{A\phi}-1\, .
\ee
Then, potential \rf{1.8} and its first derivative read, correspondingly:
\be{1.15a}
U(X)=\frac12 \left( X+1\right)^{-B/A}\left[\bar R (X+1) -f\right]
\ee
and
\ba{1.16a}
\frac{d U}{d X}= &-&\frac{B}{2A}\left( X+1\right)^{(-B/A)-1}\left[\bar R (X+1) -f\right] \nn\\&+&\frac12 \left( X+1\right)^{-B/A}\bar R\, .
\ea

To conclude this section, we would like to remind that in multidimensional case, to avoid the effective four-dimensional fundamental constant variation, it is necessary
to provide the mechanism of the internal spaces stabilization. In these models, the scale factors of the internal spaces play the role of
additional scalar fields (geometrical moduli/gravexcitons \cite{GZ(PRD1997)}).
To achieve their stabilization, an effective potential should have  minima with respect to all scalar fields (gravexcitons and scalaron).
Our previous analysis (see e.g. \cite{GZ(PRD2000)}) shows that for a model of the form \rf{1.6} the stabilization is possible only for the
case of negative minimum of the potential $U(\phi )$. However, it is not difficult to realize that it is impossible to freeze out the internal
spaces in such AdS universe. Indeed, in these models scalar fields decrease their amplitude of oscillations around a minimum during the stage of
expansion of the Universe (due to a friction term in dynamical equation of the form of \rf{5.2} below) until the Universe reaches its maximum. Then, the Universe turns to the stage of contraction and  the amplitudes of scalar fields start to increase again. Thus, geometrical moduli are not stabilized in such models. Therefore, in our
present paper we do not investigate the problem of the extra dimension stabilization but we focus our attention on the dynamics of the scalaron
field and the Universe in four-dimensional case.

\section{The $R^2+R^4$-model\label{model}}
\setcounter{equation}{0}

In this section we analyze a model with
curvature-quadratic and curvature-quartic correction terms of the
type
\be{4.1}
f(\bar{R})=\bar{R}+\alpha\bar{R}^{2}+\gamma\bar{R}^{4}-2\Lambda_{D}\,
. \ee
We start our investigation for an arbitrary number of dimensions
$D$ but in the most particular examples  we shall put $D=4$ (unless stated otherwise). First of all, we define the relation between the
scalar curvature $\bar{R}$ and the scalaron field $\phi$.
According to eq. \rf{1.7} we have:
\be{4.2}
f' = e^{A\phi} = 1 +2\alpha \bar{R} + 4\gamma
\bar{R}^3\; .
\ee
The definition \rf{1.7} $f'=\exp(A\phi)$ clearly indicates that we choose the positive branch $f'>0$. For our model \rf{4.1}, the surfaces $f'=0$
as a functions
$\bar R = \bar R(\alpha , \gamma)$ are given in Fig. \rf{f=0}. As it easily follows from Eq. \rf{4.2}, points where all three values $\bar R, \gamma$ and
$\alpha$ are positive correspond to the region $f' >0$. Thus, this picture shows that we have one simply connected region $f'>0$ and two disconnected regions $f'<0$.
\begin{figure}
  \center
    \includegraphics[width=3.in,height=2.2in]{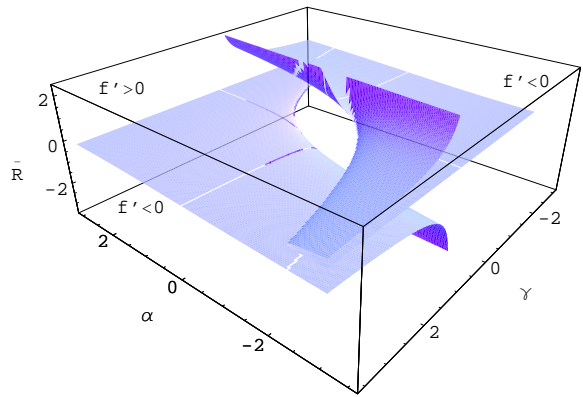}\\
\caption {The surfaces $f'=0$ as a functions
$\bar R = \bar R(\alpha , \gamma)$ for the model \rf{4.1}.\label{f=0}}
\end{figure}

Eq. \rf{4.2} can be rewritten equivalently in the form:
\be{s1} \bar R^3+\frac{\alpha}{2\gamma}\bar R-\frac
1{4\gamma}X=0,\ee
\ba{4.4} X \equiv e^{A\phi} - 1\, ,\quad &&-\infty <\phi
<+\infty\nn\\ \Longrightarrow &&-1<X<+\infty\, . \ea
Eq. \rf{s1} has three solutions $\bar R_{1,2,3}$, where one or three of
them are real-valued. Let
\be{s4}
q:=\frac{\alpha}{6\gamma},\quad r:=\frac 1{8\gamma}X.
\ee
The sign of the discriminant
 \be{s5}
 Q:=r^2+q^3
 \ee
 defines the
number of real solutions:
\ba{s6}
Q>0&\qquad
\Longrightarrow \qquad & \Im \bar R_1=0,\quad \Im \bar R_{2,3}\neq 0\, ,\nn\\
Q=0&\qquad \Longrightarrow \qquad & \Im \bar R_i=0 \ \forall i,
\quad \bar R_1=\bar R_2\, ,\nn \\
Q<0&\qquad \Longrightarrow \qquad & \Im \bar R_i=0 \ \forall i\, .
\ea
Physical scalar curvatures correspond to real solutions $\bar
R_i(X)$.
It is the most convenient to consider $\bar R_i=\bar R_i(X)$ as
solution family depending on the two additional parameters
$(\alpha,\gamma)$: $sign(\alpha)=sign(\gamma)\;\Longrightarrow
Q>0$, $sign(\alpha)\neq sign(\gamma)\;\Longrightarrow Q\gtrless
0$. The case $sign(\alpha)=sign(\gamma)$ was considered in our paper \cite{SZGC2006}.
In the present paper we investigate the most interesting case $sign(\alpha)\neq sign(\gamma)$ of multi-valued solutions.

For $Q>0$ the single real solution $\bar R_1$ is given
as
\be{s7}\bar
R_1=\left[r+Q^{1/2}\right]^{1/3}+\left[r-Q^{1/2}\right]^{1/3} := z_1+z_2\, ,
\ee
where we can define $z_{1,2}$ in the form:
\ba{4.5}
&&z_{1,2}^3=p\, e^{\pm\theta}\, , \quad p^2 = r^2-Q=-q^3\, ,\nn\\&&\cosh (\theta ) = \frac{r}{\sqrt{-q^3}}\, .
\ea
Taking into account eq. \rf{s4}, the function $X$ reads
\be{4.6}
X(\theta ) = 8\gamma \sqrt{-q^3}\cosh (\theta)\, .
\ee

 The three real solutions $\bar R_{1,2,3}(X)$ for $Q<0$ are
given as
\ba{s8}
\bar R_1&=&s_1+s_2,\nn\\
\bar R_2&=&\frac 12 (-1+i\sqrt 3)s_1+\frac 12 (-1-i\sqrt 3)s_2\nn
\\&=&e^{i\frac{2\pi}3}s_1+e^{-i\frac{2\pi}3}s_2,\nn\\
\bar R_3&=&\frac 12 (-1-i\sqrt 3)s_1+\frac 12 (-1+i\sqrt 3)s_2\nn
\\&=&e^{-i\frac{2\pi}3}s_1+e^{i\frac{2\pi}3}s_2,
 \ea
  where we can
fix the Riemann sheet of $Q^{1/2}$ by setting in the definitions
of $s_{1,2}$
 \be{s9} s_{1,2}:=\left[r\pm i|Q|^{1/2}\right]^{1/3}.
\ee
 A simple Mathematica calculation gives for Vieta's relations
from \rf{s8}
\ba{s10}
\bar R_1+\bar R_2+\bar R_3&=&0,\nn\\
\bar R_1\bar R_2+\bar R_1\bar R_3+\bar R_2\bar R_3&=&-3s_1s_2=3q,\nn\\
\bar R_1\bar R_2\bar R_3&=&s_1^3+s_2^3=2r.
\ea
 In order to work
with explicitly real-valued $\bar R_i$ we rewrite $s_{1,2}$ from
\rf{s9} as follows
\ba{s11} s_{1,2}&=&|b|^{1/3}e^{\pm i\vartheta/3},\nn\\
|b|^2&=&r^2+|Q|=r^2-Q=-q^3, \nn\\
\cos(\vartheta)&=&\frac{r}{|b|}=\frac {r}{\sqrt{-q^3}}.
\ea
 and get via
\rf{s8}
\ba{s12}
\bar R_1&=&s_1+s_2=2|b|^{1/3}\cos (\vartheta/3),\\
\bar R_2&=&e^{i\frac{2\pi}3}s_1+e^{-i\frac{2\pi}3}s_2=2|b|^{1/3}\cos \left(\vartheta/3+2\pi/3\right),\nn\\
\bar
R_3&=&e^{-i\frac{2\pi}3}s_1+e^{i\frac{2\pi}3}s_2=2|b|^{1/3}\cos
\left(\vartheta/3-2\pi/3\right)\nn
\ea
or
\ba{s13}
\bar R{_k}&=&2|b|^{1/3}\cos \left(\frac{\vartheta+2\pi
k}3\right)\nn\\&=&2\sqrt{-q}\cos \left(\frac{\vartheta+2\pi k}3\right),\;
k=-1,0,1.\quad
\ea
   In order to understand the qualitative
behavior of these three real-valued solutions as part of the
global solution picture we first note that, according to \rf{s1},
we may interpret $X$ as single-valued function
\be{s14}
X(\bar R)=4\gamma \bar R^3+2\alpha \bar R
\ee
 and look
what is happening when we change $(\alpha,\gamma)$. Obviously, the
inverse function $\bar R(X)$ has three real-valued branches when
$X(\bar R)$ is not a monotonic function but instead has a minimum
and a maximum, i.e. when
\be{s15}
\partial_{\bar R}X:= X'=12 \gamma \bar R^2+2\alpha=0\;
\Longrightarrow \; \bar R^2=-\frac{\alpha}{6\gamma}
\ee
has two real solutions $\bar R_\pm=\pm \sqrt{-\alpha/(6\gamma)}$
and corresponding extrema\footnote{It is worth of noting that $ f''(\bar R_\pm) =X'(\bar R_\pm) =0$.}
\be{s15a}
X(\bar R_\pm)=\frac 43 \alpha \bar R_\pm\, .
\ee
It should
hold $\sign (\alpha)\neq \sign (\gamma)$ in this case so that we
find
\ba{s15b}
\gamma>0,\alpha<0:&\;& X_{max}=X(\bar R_-), \
X_{min}=X(\bar
R_+)\nn\\
\gamma<0,\alpha>0:&\; & X_{max}=X(\bar R_+), \ X_{min}=X(\bar
R_-)\,.\nn\\
\ea
 The transition from the three-real-solution regime to
the one-real-solution regime occurs when maximum and minimum
coalesce at the inflection point
\be{s16}
\bar R_+=\bar R_-=0\; \Longrightarrow \;
\alpha=0, \ \gamma \neq 0\, .
\ee
 (We note that here we consider the
non-degenerate case $\gamma\neq 0$. Models with $\gamma=0$ are
degenerate ones and are characterized by quadratic scalar
curvature terms only.) Due to $-1\le X\le +\infty$ we may consider
the limit $X\to +\infty$ where in leading approximation
\be{s17}
4\gamma \bar R^3\approx X\to +\infty \ee so that \be{s18}
\bar R(X\to \infty)\to \sign(\gamma)\times \infty\, .
\ee
Leaving
the restriction $X\ge -1$ for a moment aside, we have found that
for $\alpha\gamma<0$
there exist three real solution branches $\overline{\mathcal{R}}_{1,2,3}$:
\ba{s19} \gamma>0:&\; & -\infty \le \overline{\mathcal{R}}_1\le \bar R_-, \quad -\infty \le X \le X_{max},\nn\\
&& \bar R_-\le \overline{\mathcal{R}}_2 \le \bar R_+, \quad X_{max}\ge X \ge X_{min},\nn\\
&& \bar R_+\le \overline{\mathcal{R}}_3\le +\infty, \quad X_{min}\le X \le +\infty, \nn\\
\gamma<0:&\; & -\infty \le \overline{\mathcal{R}}_1 \le \bar R_-, \quad +\infty\ge X\ge X_{min},\nn\\
&&\bar R_-\le \overline{\mathcal{R}}_2 \le \bar R_+, \quad X_{min}\le X \le X_{max},\nn\\
&&\bar R_+\le \overline{\mathcal{R}}_3 \le +\infty, \quad X_{max}\ge X \ge
-\infty.\nn\\\ea
It remains for each of these branches to check which
of the solutions $\bar R_k$ from \rf{s13} can be fit into this
scheme. Finally, one will have to set the additional restriction
$X\ge -1$ on the whole picture.

\section{The fitting procedure}
\setcounter{equation}{0}

 We start by considering a concrete example. For
definiteness, let us assume $\gamma>0,\alpha<0$. The pairwise
fitting of the various solution branches should be performed at
points where $Q=0$ and different branches of the three-solution
sector are fitted with each other or to the branches of the
one-solution sector. The points with $Q=0$ correspond to the
$X-$values $X_{min}$ and $X_{max}$. Explicitly we have from
\rf{s15a}
\be{f1} X(\bar R_\pm)=\frac 43\alpha\bar R_\pm=\pm\frac
43\alpha\sqrt{-\frac{\alpha}{6\gamma}} \ee
 and for the concrete
configuration $\gamma>0,\alpha<0$
\ba{f2} X_{max}&=&X(\bar R_-)=-\frac
43\alpha\sqrt{-\frac{\alpha}{6\gamma}}\ge 0, \; \nn\\X_{min}&=&X(\bar
R_+)=\frac 43\alpha\sqrt{-\frac{\alpha}{6\gamma}}\le 0\,.\ea
 Next,
we find from the defining equation \rf{s11} for the angle
$\vartheta$ that at $Q=0$ it holds
\be{f3} \cos (\vartheta)=\frac{r}{|b|}=\frac{r}{|r|} \ee
so that
\ba{f4}  X_{max}\ge 0 &\Longrightarrow& r>0 \;
\Longrightarrow \; \cos(\vartheta)=1 \; \nn\\&\Longrightarrow&
\vartheta=2\pi m,\ m\in \ZZ\,,\nn\\
X_{min}\le 0 &\Longrightarrow& r<0 \;
\Longrightarrow \;\cos(\vartheta)=-1 \;\nn\\&\Longrightarrow&
 \vartheta=\pi+2\pi n,\ n\in \ZZ\,. \ea
   Now, the fitting of
the various solution branches can be performed as follows (see Fig. \rf{zigzag}).
\begin{figure}
  \center
    \includegraphics[width=3.2in,height=2.3in]{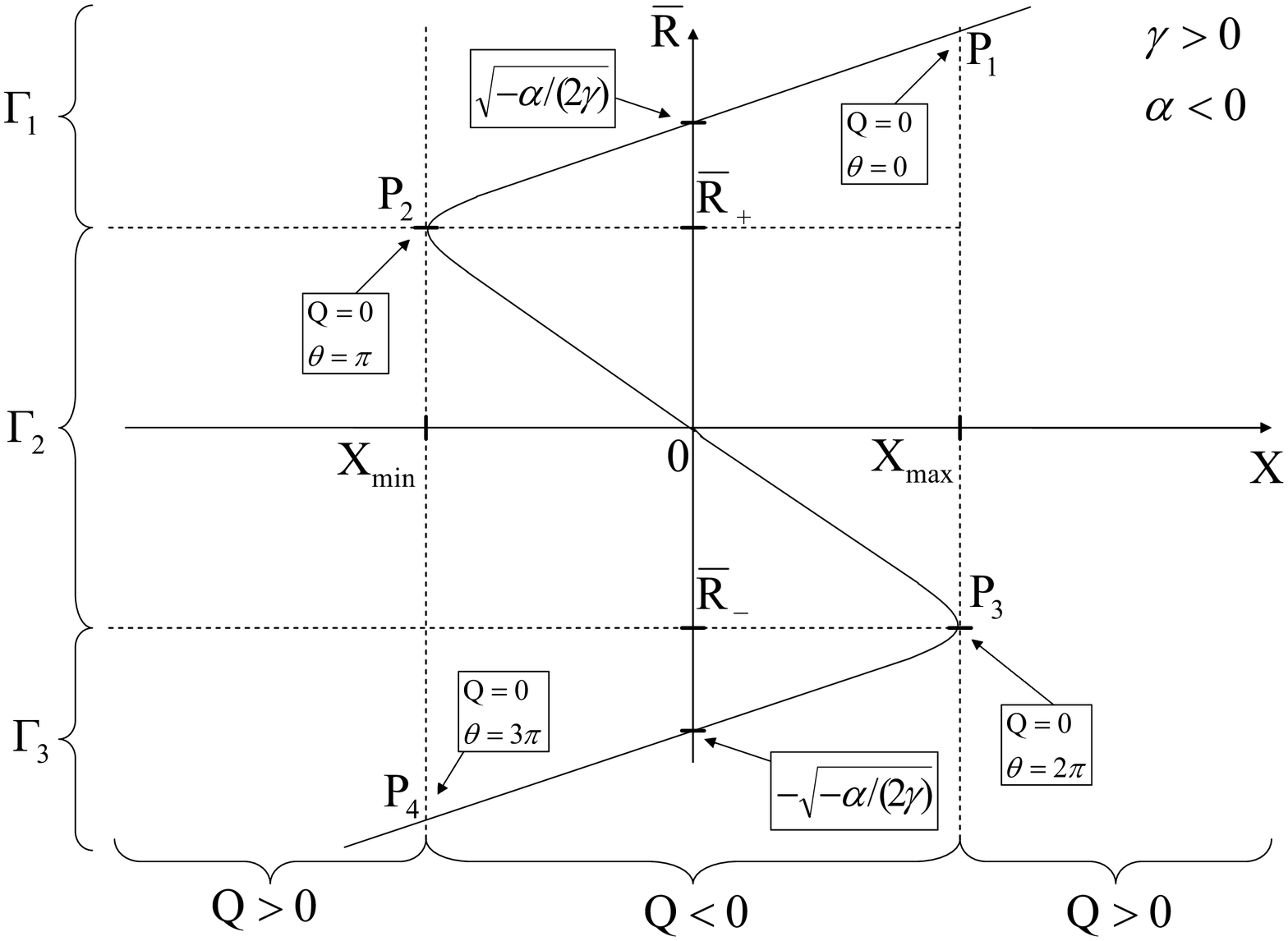}\\
\caption {The schematic drawing of the  real solution branches
and the matching points $P_{1,2,3,4}$ This figure shows that points $P_{2,3}$ (correspondingly, $\theta=\pi,2\pi$) and points
$P_{1,4}$ (correspondingly, $\theta=0,3\pi$) have different nature. So,  $P_{2,3}$ and  $P_{1,4}$ we shall call branching points and
monotonic points (in the sense that function $\bar R$ is monotonic in the vicinity of these points), respectively.\label{zigzag}}
\end{figure}
We
start with the branch $\Gamma_1:=(R_+\le \bar R\le
+\infty,X_{min}<X<+\infty)$. Moving in on this branch from
$X\approx +\infty$ we are working in the one-solution sector $Q>0$
with
\be{f5} \bar
R(\Gamma_1;Q)=\left[r+Q^{1/2}\right]^{1/3}+\left[r-Q^{1/2}\right]^{1/3}\ee
 until we hit $Q=0$ at $X=X_{max}$. At this point
$P_1:=(\Gamma_1,X=X_{max})\in \Gamma_1$ we have to perform the
first fitting. Due to $r>0$ we may choose
\be{f6} \bar R(\Gamma_1;Q=0)=2r^{1/3}=2|b|^{1/3}\ee
 so that as
simplest parameter choice in \rf{s13} we set
\be{f7} P_1=(\Gamma_1,X=X_{max},Q=0)\ \mapsto \ \vartheta=0,\
k=0\,. \ee
 Hence, the parametrization for $(\Gamma_1,Q<0)$ will be
given as
\be{f8} \bar R(\Gamma_1,Q<0)=2|b|^{1/3}\cos (\theta/3). \ee
 For
later convenience, we have replaced here the $\vartheta$ from the
equations \rf{s12}, \rf{s13} by
$\theta$. The reason will become clear from the subsequent
discussion. We note that on this $\Gamma_1-$segment we may set
$\vartheta=\theta$. Let us further move on $\Gamma_1$ until its
end at $X_{min}$, where again $Q=0$. Because there was no other
point with $Q=0$ on this path, the smoothly changing $\theta$ can
at this local minimum $P_2=\Gamma_1\cap \Gamma_2=(X=X_{min},\bar
R=\bar R_+)$ only take one of the values $\theta=\pm \pi$. For
definiteness we choose it as $\theta (P_2)=\pi$. Hence, it holds
\ba{f9} \bar R(P_2)&=&2|b|^{1/3}\cos
(\pi/3)=|b|^{1/3}\nn\\&=&\sqrt{-q}=\sqrt{-\alpha/(6\gamma)}=\bar R_+\qquad\ea
as it should hold. For convenience, we may parameterize our
movement on the cubic curve by simply further increasing $\theta$.
This gives for moving on $\Gamma_2=(\bar R_+\ge \bar R\ge \bar
R_-,X_{min}\le X \le X_{max})$ from the local minimum at $P_2$ to
the local maximum at $P_3=\Gamma_2\cap \Gamma_3=(X=X_{max},\bar
R=\bar R_-)$ a further increase of $\theta$ by $\pi$ up to $\theta
(P_3)=2\pi$. Accordingly, we find the complete consistency
\ba{f10} \bar R(P_3)&=&2|b|^{1/3}\cos
(2\pi/3)=-|b|^{1/3}\nn\\&=&-\sqrt{-q}=-\sqrt{-\alpha/(6\gamma)}=\bar
R_-\,. \ea
By further increasing $\theta$ up to $\theta=3\pi$ we
reach the point $P_4=(X=X_{min},Q=0)\in \Gamma_3$ with \be{f11}
\bar R(P_4)=2|b|^{1/3}\cos (3\pi/3)=-2|b|^{1/3}=-2|r|^{1/3}.
\ee
Because of $r<0$ we can smoothly fit it to the one-solution branch
\be{f12} \bar
R(\Gamma_4,Q)=\left[r+Q^{1/2}\right]^{1/3}+\left[r-Q^{1/2}\right]^{1/3}
\ee
 by setting trivially
 \be{f13} \bar
R(P_4)=2(-|r|)^{1/3}=-2|r|^{1/3}.
\ee
  Summarizing, we arrived at
a very simple and transparent branch fitting picture, where all
the movement in the three-solution sector can be parameterized by
choosing the effective angle as $\theta \in [0,3\pi]$. Finally, we
have to fit this picture in terms of smoothly varying
$\theta\in[0,3\pi]$ with the three-solutions $\bar R_k$ from
\rf{s13}. For this purpose we note that the single value
$\vartheta\in [0,\pi]$ in \rf{s13} is a projection of our smoothly
varying $\theta\in[0,3\pi]$. Fixing an arbitrary $\vartheta$ one
easily finds the following correspondences
\be{f14} \theta(\Gamma_1,\vartheta)=\vartheta,\;
\theta(\Gamma_2,\vartheta)=2\pi-\vartheta,\;
\theta(\Gamma_3,\vartheta)=2\pi+\vartheta
\ee
 and hence
\ba{f15}
\bar R[\theta(\Gamma_1,\vartheta)]&=&2|b|^{1/3}\cos\left(\frac \vartheta 3\right)=
\bar{R}_{(k=0)}=\overline{\mathcal{R}}_{3}\;,\nn\\
\bar
R[\theta(\Gamma_2,\vartheta)]&=&2|b|^{1/3}\cos\left(\frac{2\pi-
\vartheta} 3\right)\nn\\&=&
2|b|^{1/3}\cos\left(\frac{\vartheta-2\pi} 3\right)=\bar R_{(k=-1)}=\overline{\mathcal{R}}_{2},\nn\\
\bar
R[\theta(\Gamma_3,\vartheta)]&=&2|b|^{1/3}\cos\left(\frac{2\pi+\vartheta}3\right)\nn\\&=&\bar
R_{(k=1)}=\overline{\mathcal{R}}_{1}\,. \ea
Analogically, we can obtain rules for fitting procedure in the
case $\gamma<0,\alpha>0$.
So, all the fitting mechanism is clear now and
can be used in further considerations.

\section{Dynamics of the Universe and scalaron}
\setcounter{equation}{0}

To study the dynamics of the Universe in our model, we assume that the four-dimensional metric $g$ in \rf{1.9} is
spatially flat Friedmann-Robertson-Walker one:
\be{5.1}
g=-dt\otimes dt + a^2(t)d\vec{x}\otimes d\vec{x}\, .
\ee
Thus, scalar curvatures
$R$ and $\bar R$ and scalaron $\phi$ are functions of time. Therefore, Eq. \rf{1.11a} for homogeneous field $\phi$ reads
\be{5.2}
\ddot{\phi}+3H\dot{\phi}+\frac{d U}{d\phi}=0\, ,
\ee
where the Hubble parameter $H=\dot{a}/a$ and the dotes denote the differentiation with respect to time $t$. Potential $U$ is defined by
Eq. \rf{1.8}. Because $U$ depends on $\bar R$ which is a multi-valued function of $\phi$ (or, equivalently, of $X$), the potential $U$ is
also a multi-valued function of $X$ (see Fig. \rf{U(X)})\footnote{In spite of the divergency of $d\bar R/d X$ in the branching points $P_{2,3}$, the derivatives $dU/dX$ are finite in these points in accordance with  Eq. \rf{1.16a}.  Moreover, $\bar R$ and $X$ have the same values in branching points for different branches. Therefore, the branches arrive at the branching points with the same values of $dU/dX$ and Fig. \rf{U(X)} clearly shows it.
}.
\begin{figure}
  \center
    \includegraphics[width=3.3in,height=2.1in]{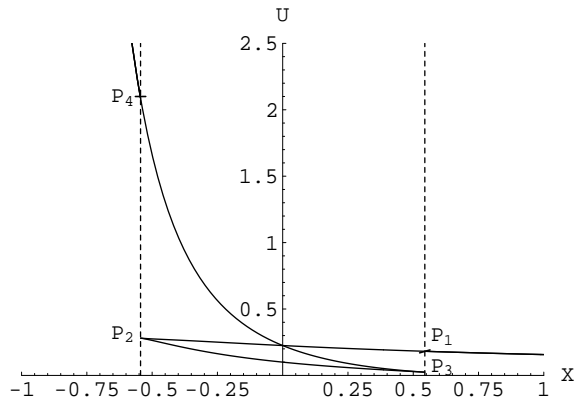}\\
\caption {The form of the potential \rf{1.8} as a multi-valued function of $X=e^{A\phi}-1$ in the case
$D=4$, $\Lambda_4=0.1$, $\gamma =1$ and $\alpha =-1$. Points $P_{1,2,3,4}$ are defined in Fig. \rf{zigzag}.\label{U(X)}}
\end{figure}
However, our previous analysis shows that we can avoid this problem making $X$ and $\bar R$ single-valued
functions of a new field $\theta$ (we would remind that we consider the particular case of $\alpha \gamma <0$ when $\alpha <0,\, \gamma >0$):
\be{5.3}
X(\theta) =
\left\{\begin{array}{ccc}
\sqrt{(1/\gamma )\left(2|\alpha |/3\right)^3}\; \cosh (\theta ) \qquad\,,\theta <0&;&\; \\
\sqrt{(1/\gamma )\left(2|\alpha |/3\right)^3}\; \cos (\theta ) \,, 0\leq\theta \leq 3\pi&;&\;\\
-\sqrt{(1/\gamma )\left(2|\alpha |/3\right)^3}\; \cosh (\theta -3\pi ) \,, \theta > 3\pi &;&\;  \\
\end{array}\right.
\ee
and
\be{5.4}
\bar R(\theta) =
\left\{\begin{array}{ccc}
2\sqrt{|\alpha|/(6\gamma)}\; \cosh (\theta /3 ) \,,\quad\qquad \theta <0&;&\; \\
2\sqrt{|\alpha|/(6\gamma)}\; \cos (\theta /3) \,, \quad 0\leq\theta \leq 3\pi&;&\;\\
-2\sqrt{|\alpha|/(6\gamma)}\; \cosh [(\theta /3) -\pi ] \,,\quad  \theta > 3\pi &.&\;  \\
\end{array}\right.
\ee
The function $X = X(\theta)$ is schematically given in Fig. \rf{X(theta)}.
\begin{figure}
  \center
    \includegraphics[width=3.1in,height=2in]{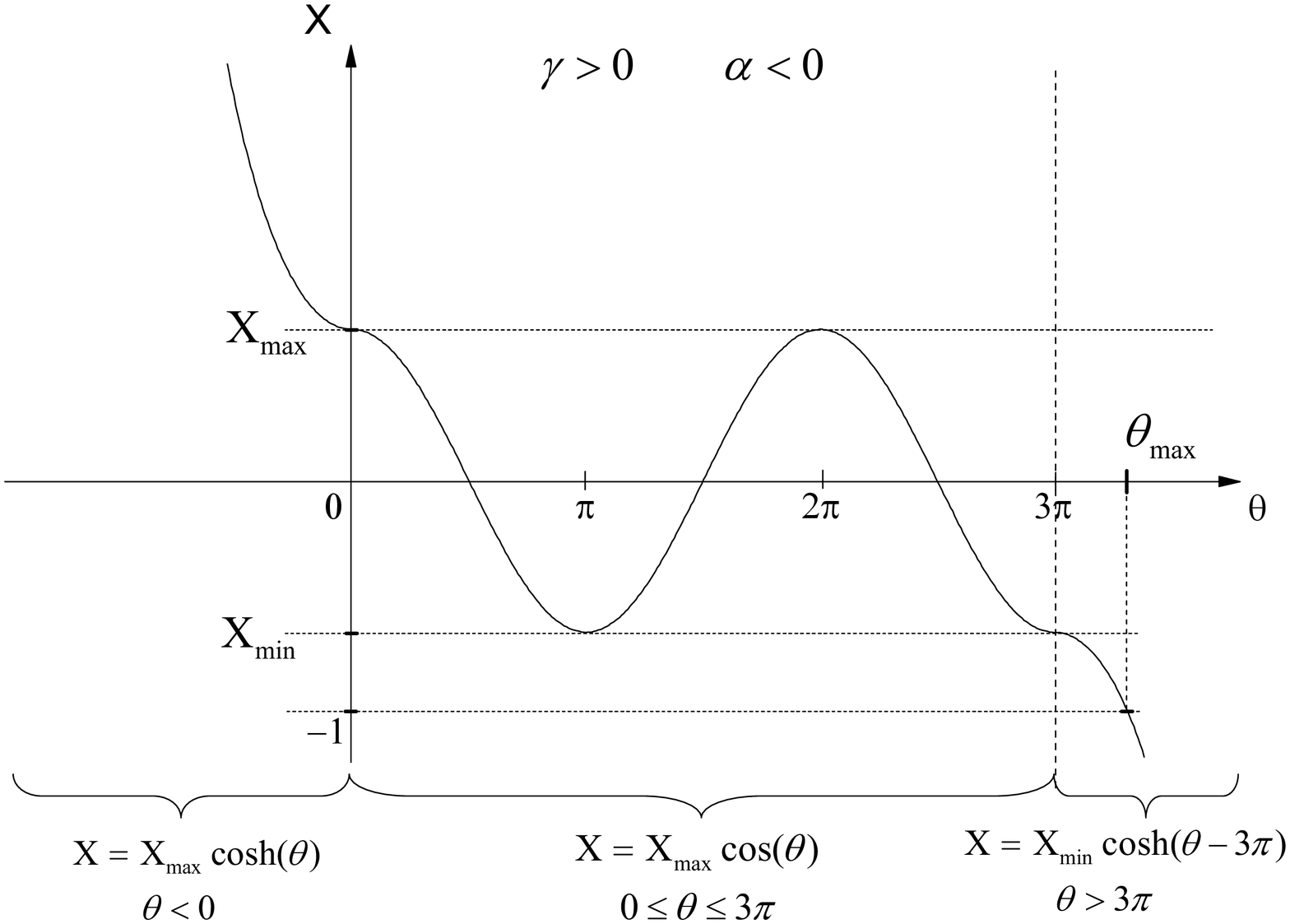}\\
\caption {The schematic drawing of Eq. \rf{5.3} in the case $X_{min}>-1$. Here, $X_{max} =\sqrt{(1/\gamma )\left(2|\alpha |/3\right)^3}\, ,\;
X_{min}=-\sqrt{(1/\gamma )\left(2|\alpha |/3\right)^3}$ and $\theta_{max}$ is defined by Eq. \rf{5.6}.\label{X(theta)}}
\end{figure}
It is necessary to keep in mind that we consider the case
$f'>0 \rightarrow X>-1$. If we demand that $X_{min} >-1$ (in opposite case our graphic $X(\theta)$ will be cut into two disconnected parts) then
the parameters $\alpha$ and $\gamma$ should satisfy the inequality:
\be{5.5}
X_{min}=-\sqrt{(1/\gamma )\left(2|\alpha |/3\right)^3} >-1 \; \Rightarrow \;|\alpha| \leq \frac32 \gamma^{1/3}\, .
\ee
The maximal value of $\theta $ (which is greater than $3\pi$ in the case $X_{min} >-1$) is defined from the transcendental equation
\ba{5.6}
&&\frac12\left[\left(c+\sqrt{c^2-1}\, \right)^{1/3}+ \left(c-\sqrt{c^2-1}\, \right)^{1/3}\right]=\nn\\ &&= \cosh [(\theta_{max} /3) -\pi ]\, ,
\ea
where $c := \left(\sqrt{(1/\gamma )\left(2|\alpha |/3\right)^3}\right)^{-1} $. The limit $X\to -1$ corresponds to the limit $\theta \to \theta_{max}$.
With the help of Eq. \rf{5.3} and formula
\be{5.7}
\frac{d \phi}{d \theta} = \frac{1}{A(X+1)}\frac{d X}{d \theta}
\ee
we can also get the following useful expressions:
\be{5.8}
\left. \frac{d \phi}{d \theta}\right|_{\theta =0,\pi,2\pi,3\pi}  = \left.\frac{d X}{d \theta}\right|_{\theta =0,\pi,2\pi,3\pi} =0\, .
\ee

It can be easily verified that field $\theta$ satisfies the equation:
\be{5.9}
\ddot \theta +3 H\dot \theta + \Gamma (\theta){\dot \theta}^2 + G (\theta )\frac{d U}{d \theta} =0\, .
\ee
Here, we introduce the one dimensional metric on the moduli space $G (\theta )\equiv G^{11} = (G_{11})^{-1}=(d\phi /d\theta)^{-2}$ with the
corresponding Christoffel symbol $\Gamma (\theta) \equiv \Gamma^{1}_{11}=(1/2)G^{11}(G_{11})_{,\, \theta} = (d^2\phi /d\theta^2)/(d\phi /d\theta)$.

\subsection{Properties of the potential $U(\theta )$}

As we mentioned above, the potential \rf{1.8} as a function of $\theta $ is a single-valued one.
Now, we want to investigate analytically some general properties of $U(\theta )$. In this subsection, $D$ is an arbitrary number of dimensions and signs of
$\alpha$ and $\gamma$ are not fixed if it is not specified particularly.

First, we consider the {\it extrema of the potential} $U(\theta )$. To find the extremum points we solve the equation
\be{5.10}
\frac{d U}{d\theta} = \frac{dU}{d\phi}\frac{d\phi}{d\theta}=\frac{dU}{d\phi}\frac{1}{A(X+1)}\frac{dX}{d\theta}=0\, .
\ee
Therefore, the extrema correspond either to the solutions of the equation \rf{1.12a} $dU/d\phi =0$ for finite $dX/d\theta$ ($X>-1$) or
to the solutions of the equation $dX/d\theta=0$ ($X >-1$) for finite $dU/d\phi$.  The form of the potential $U$ ( see Eqs. \rf{1.8a} and \rf{1.8})
shows that this potential and its derivative $dU/d\phi$ is finite for $X>-1$. Thus, as it follows from Eq. \rf{5.8}, the potential $U(\theta )$ has extrema at the matching points $\theta =0,\pi,2\pi,3\pi$.
Additional extremum points are real solutions of the equation \rf{1.12a}. For our model \rf{4.1} this
equation reads:
\ba{5.11}
&&\bar R^{4}\gamma\left(\frac{D}{2}-4\right)+\bar
R^{2}\alpha\left(\frac{D}{2}-2\right)\nn\\&+&\bar
R\left(\frac{D}{2}-1\right)-D\Lambda_{D} =0\;.
\ea
The form of this equation shows that there are two particular cases: $D=8$ and $D=4$. The $D=8$ case was considered in \cite{SZGC2006}. Let us consider now the
case $D=4$:
\be{5.12}
 \bar R^4 -\frac{1}{2\gamma}\bar R +\frac{2\Lambda_4}{\gamma}=0\, .
\ee
It is worth of noting that parameter $\alpha$ disappeared from this equation. Thus $\alpha$ has no an influence on a number
of additional extremum points.
To solve this quartic equation, we should consider an auxiliary cubic equation
\be{5.13}
u^3-\frac{8\Lambda_4}{\gamma}u-\frac{1}{4\gamma^2}=0\, .
\ee
The analysis of this equation can be performed in similar manner as we did it for the cubic equation \rf{s1}.
Let us introduce the notations:
\ba{5.14}
&&\bar q := -\frac{8}{3}\frac{\Lambda_4}{\gamma}\; ,\bar r := \frac{1}{8}\frac{1}{\gamma^2}\, ,
\nn\\&&\bar Q:= \bar r^2 +\bar q^3 = \frac{1}{\gamma^4}\left(\frac{1}{8^2}-\gamma\left(\frac{8\Lambda_4}{3}\right)^3\right).
\ea
It make sense to consider two separate cases.

1. $\sign \gamma = -\sign \Lambda_4 \quad \Rightarrow \quad \bar Q>0\, .$

In this case we have only one real solution of Eq. \rf{5.13}:
\be{5.15}
u_1 = \left[\bar r +\bar Q^{1/2}\right]^{1/3} + \left[\bar r -\bar Q^{1/2}\right]^{1/3} >0\, .
\ee
Then, solutions of the quartic \rf{5.12} are the real roots of two quadratic equations
\be{5.16}
\bar R^2 \pm \sqrt{u_1}\bar R +\frac12\left(u_1\pm\epsilon\sqrt{u_1^2+3q} \right)=0\, ,\quad \epsilon = \sign \gamma\, .
\ee
Simple analysis shows that for any sign of $\gamma$ we obtain two real solutions:
\be{5.17}
\begin{array}{ccc}
\gamma <0  \Rightarrow  \bar R^{(+)}_{1,2}=-\frac{1}{2}u_1^{1/2}\pm \sqrt{-\frac{1}{4}u_1+\frac{1}{2}(u_1^2+3q)^{1/2}}\,,\\
\phantom{\int}\\
\gamma >0 \Rightarrow
\bar R^{(-)}_{1,2}=\frac{1}{2}u_1^{1/2}\pm \sqrt{-\frac{1}{4}u_1+\frac{1}{2}(u_1^2+3q)^{1/2}}\, .\\
\end{array}
\ee

2. $\sign \gamma = \sign \Lambda_4 \quad \Rightarrow \quad \bar Q\gtrless0\, .$

It is not difficult to show that in this case the real solutions of the form of \rf{5.17}
(where we should make the evident substitution $u_1^2+3q \rightarrow u_1^2-3|q|$)
takes place if
\be{5.18}
\bar Q >0 \quad \Rightarrow \quad |\gamma |^{1/3} < \frac{3}{32|\Lambda_4|}\, .
\ee

Now, we want to investigate {\it zeros of the potential} $U$. For $f'\neq 0 \Rightarrow X\neq -1$, the condition of zeros of the
potential \rf{1.8-1} is:
\be{5.19}
\bar R f'-f =0 \quad \Rightarrow \quad 3\gamma \bar R^4 +\alpha \bar R^2 +2\Lambda_D=0\, .
\ee
Therefore, zeros are defined by equation:
\be{5.20}
\bar R^2 = -\frac{\alpha}{6\gamma} \pm \left[\left(\frac{\alpha}{6\gamma}\right)^2-\frac{2\Lambda_D}{3\gamma}\right]^{1/2}\, .
\ee
Obviously, the necessary conditions for zeros are:
\be{5.21}
 \begin{array}{ccc}
\gamma > 0 \quad &\Rightarrow& \quad \Lambda_D \leq \alpha^2/(24\gamma)\, , \\
&\phantom{\int}&\\
\gamma < 0 \quad &\Rightarrow& \quad \Lambda_D \geq -\alpha^2/(24|\gamma |)\, .\\
\end{array}
\ee
Additionally, we should check that the r.h.s. of the equation \rf{5.20} is positive.

Let us consider now {\it asymptotical behavior of the potential} $U(\theta)$.  Here, we want to investigate limits $\theta \to \theta_{max}$ and $\theta \to -\infty$. In the former case we get:
\be{5.22}
\theta \to \theta_{max} \quad \Rightarrow \quad U(\theta ) \to -\sign (f(\theta_{max}))\times \infty\, .
\ee
In the latter case we obtain:
\ba{5.23}
&&\theta \to -\infty \Rightarrow \nn\\
&&U(\theta )\sim \exp\left(\frac{8-D}{D-2}\; \theta \right)\to
\left\{\begin{array}{ccc}
+\infty  \,,\;\:\qquad D>8 \;; \\
\const >0 \,, D = 8 \;; \\
+0  \,,\quad 2<D<8 \;; \\
\end{array}\right.\nn\\
\ea
where we used Eqs. \rf{5.3} and \rf{5.4}. This result coincides with conclusions of Appendix A in \cite{SZGC2006}.

To illustrate the described above properties, we draw the potential $U(\theta )$ in Fig. \rf{pot} for the following parameters:
$D=4$, $\Lambda_4=0.1$, $\gamma =1$ and $\alpha =-1$.
\begin{figure}
  \center
    \includegraphics[width=3.4in,height=2.1in]{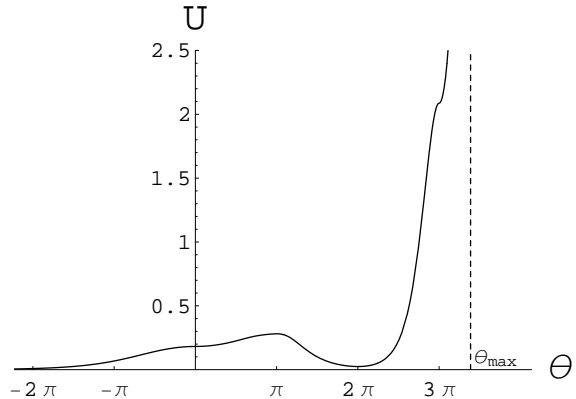}\\
\caption {The form of the potential \rf{1.8} as a function of $\theta$ in the case
$D=4$, $\Lambda_4=0.1$, $\gamma =1$ and $\alpha =-1$. For these values of the parameters, all extrema
correspond to the matching points $\theta = 0, \pi, 2\pi, 3\pi$. In the branching points $\theta =\pi,2\pi$ the potential has
local maximum and local non-zero minimum, respectively, and the monotonic points $\theta=0,3\pi$ are the inflection ones. Potential tends asymptotically to $+\infty$
when $\theta$ goes to $\theta_{max}$ and to zero when $\theta \rightarrow -\infty$.\label{pot}}
\end{figure}
These parameters contradict to the inequalities \rf{5.18} and \rf{5.21}. Therefore, $\theta = 0, \pi, 2\pi, 3\pi$ are
the only extremum points of the potential $U(\theta )$ and zeros are absent. These parameters are also satisfy the condition \rf{5.5}.
The absence of zeros means that all minima of the potential $U(\theta)$ are positive.

For our subsequent investigations, it is useful also to consider an effective force and mass squared of the field $\theta $. As it follows from Eq. \rf{5.9}, the effective force:
\be{5.24}
F= -G(\theta) \frac{dU}{d\theta}\, .
\ee
Varying Eq. \rf{5.9} with respect to field $\theta $ we obtain dynamical equation for small fluctuation $\delta \theta$ where
mass squared reads:
\be{5.25}
m^2_{\theta} = G(\theta )\frac{d^2 U}{d\theta^2}+\frac{d G(\theta)}{d\theta}\frac{dU}{d\theta }\, .
\ee
In Fig. \rf{force-mass} we show the effective force and the mass squared as functions of $\theta$ for the potential drawn in Fig. \rf{pot}.
\begin{figure*}[htbp]
\center{\includegraphics [width = 3in, height = 2.1in ]{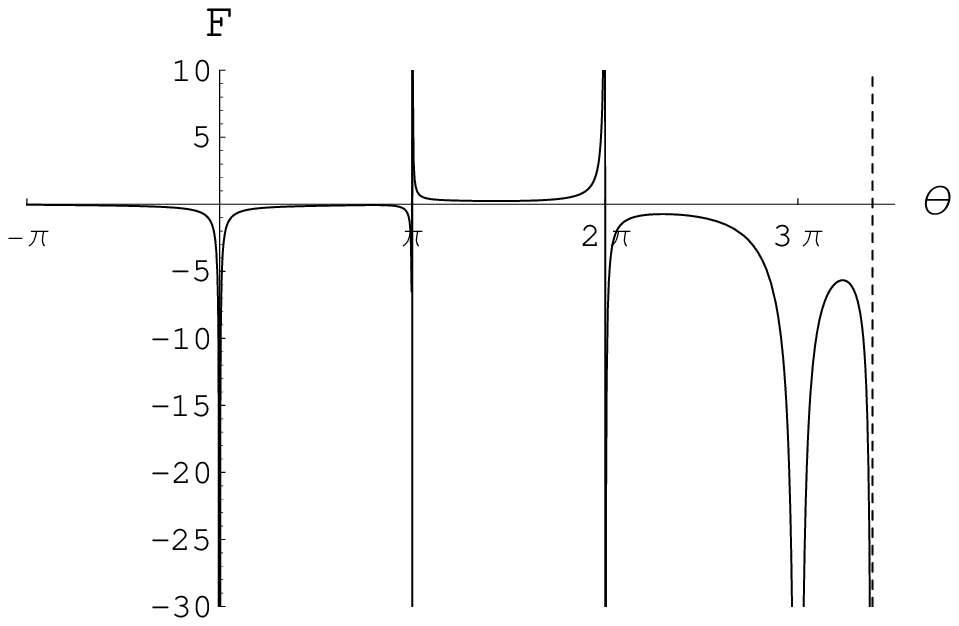}
\includegraphics [width = 3in, height = 2.1in ]{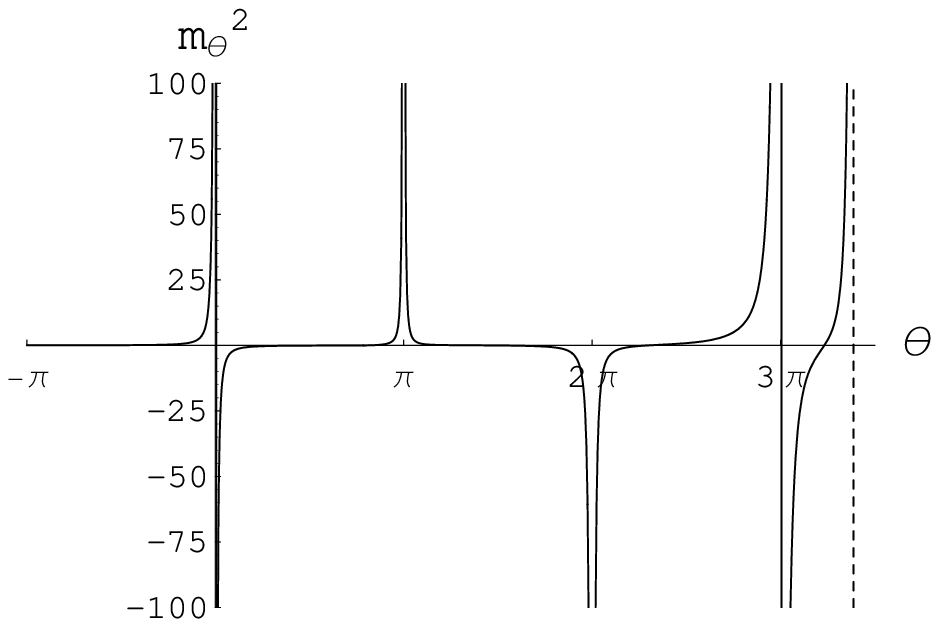}}
\caption{The effective force \rf{5.24} (left panel) and the mass squared \rf{5.25} (right panel) for the potential
$U(\theta )$ drawn in Fig. \rf{pot}. These pictures clearly show singular behavior of $F$ and $m^2_{\theta}$
in the matching points $\theta = 0, \pi, 2\pi, 3\pi$.
\label{force-mass}}
\end{figure*}
These figures
indicate that field $\theta $ may have very nontrivial behavior. This non-triviality  follows from two reasons. First, the field $\theta $ has non-canonical
kinetic term which result in appearing of non-flat moduli space metric $G(\theta )$ and derivative of $G(\theta )$ in Eq. \rf{5.9}. Second, the function $G(\theta ) $ has singular
behavior at the matching points $\theta = 0, \pi, 2\pi, 3\pi$.
Thus, our intuition does not work
when we want to predict dynamical behavior of fields with equations of the form of \rf{5.9} with potential drawn in Fig. \rf{pot}, especially when field approaches the matching points. It is necessary to solve equations analytically or to investigate them numerically.  Such analysis for our model is performed in the next subsection where
we concentrate our attention to the most interesting case  where all extrema correspond to the matching points $\theta = 0, \pi, 2\pi, 3\pi$.

\subsection{Dynamical behavior of the Universe\\ and field $\theta$}

Now, we intend to investigate dynamical behavior of scalar field $\theta$ and the
scale factor $a$ in more detail. There are no analytic
solutions for considered model. So, we use numerical calculations.
To do it, we apply a Mathematica package proposed in \cite{KP} and
adjusted to our models and notations in Appendix A of our paper \cite{SZPRD2009}. According to these notations, all dimensional quantities in our graphics
are given in the Planck units. Additionally, in the present paper we should remember that metric on the moduli space is
not flat and defined in Eq. \rf{5.9}. For example, the canonical momenta and the kinetic energy read:
\ba{5.26}
P_{\theta} &=& \frac{a^3}{\kappa_4^2}G_{11}\dot \theta = \frac{a^3}{\kappa_4^2}\left(\frac{d\phi}{d\theta}\right)^2\dot\theta\, ,\\
\nn\\E_{kin} &=& \frac{1}{2\kappa_4^2} G_{11}\dot \theta^2=\frac{\kappa_4^2}{2a^6}G^{11}P_{\theta}^2 = \frac{1}{2\kappa_4^2} \left(\frac{d\phi}{d\theta}\right)^2\dot\theta^2\,,\nn
\ea
where $8\pi G \equiv \kappa_4^2$ and $G$ is four-dimensional Newton constant. To understand the dynamics of the Universe, we shall also draw the
Hubble parameter:
\be{5.27}
3 \left(\frac{\dot a}{a}\right)^2 \equiv 3 H^2 =
\frac12 G_{11} \dot \theta^2 +U(\theta)\,
\ee
and the acceleration parameter:
\be{5.28}
q \equiv \frac{\ddot a}{H^2 a} = \frac{1}{6H^2} \left(-4 \times
\frac12 G_{11} \dot \theta^2 + 2U(\theta)\right)\, .
\ee

Fig. \rf{force-mass} shows that the effective force changes its sign and the mass squared preserves the sign when $\theta$ crosses the branching points
$\pi, 2\pi$ and vise verse, the effective force preserves the sign and the mass squared changes the sign when $\theta$ crosses the monotonic points
$0, 3\pi$. Therefore, it make sense to consider these cases separately.

\subsubsection{branching points
$\theta =\pi, 2\pi$}

\begin{figure*}[htbp]
\centerline{\includegraphics[width=3.0in,height=2.0in]{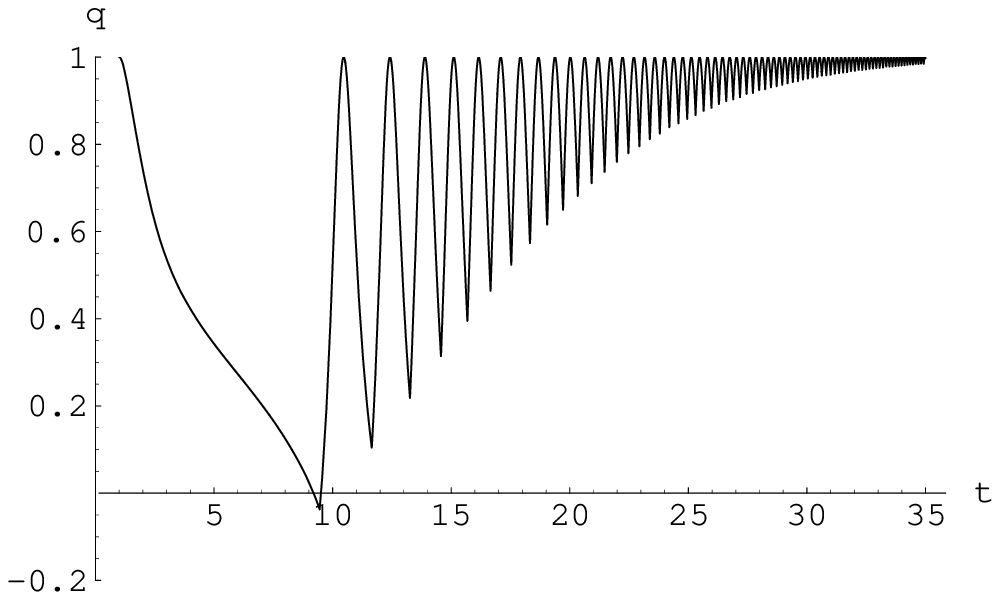}
\includegraphics[width=3.0in,height=2.0in]{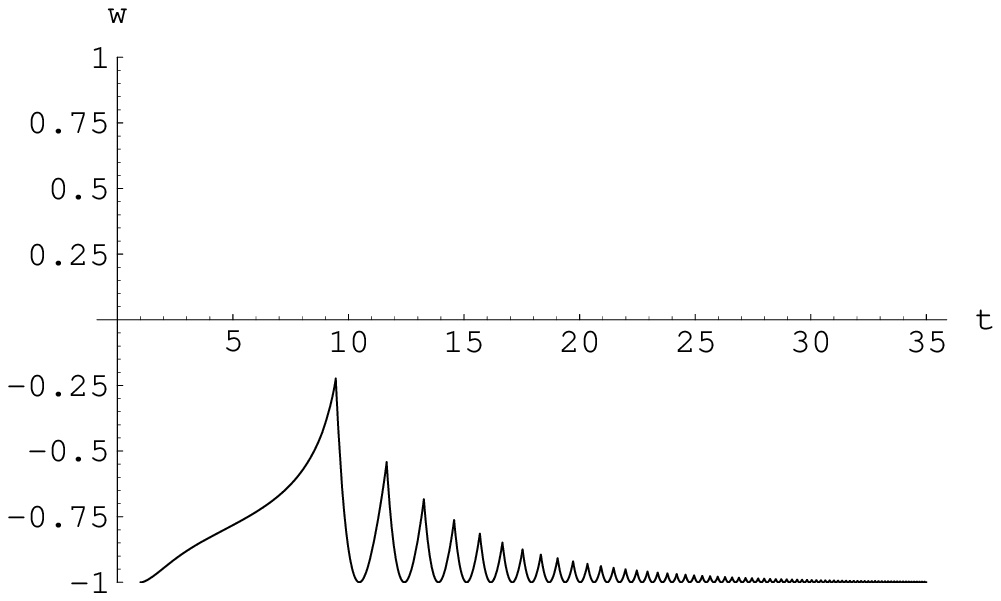}}
\caption {The parameter of acceleration $q(t)$ (left panel) and the equation of state parameter $\omega (t)$ (right panel) for the scale factor in Fig. \rf{H,a}.
\label{q}}
\end{figure*}

First, we consider the dynamical behavior of the Universe and scalaron in the vicinity of the branching point $\theta =2\pi$ which is the local minimum
of the potential in Fig. \rf{pot}. The time evolution of scalaron field $\theta$ and its kinetic energy $E_{kin}$ are drawn in Fig. \rf{theta}.
Here and in all pictures below, we use the same parameters as in Fig. \rf{pot}. The time $t$ is measured in the Planck times and classical evolution starts at $t=1$.
For the initial value of $\theta$ we take $\theta_{initial}=3.5$.
\begin{figure*}[htbp]
\centerline{\includegraphics[width=3.0in,height=1.9in]{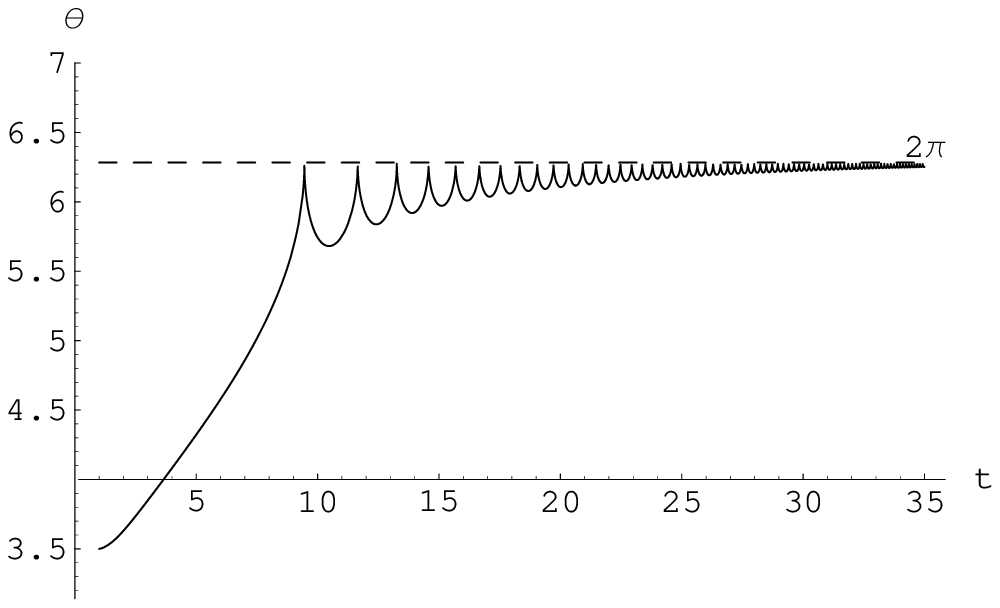}
\includegraphics[width=3.0in,height=1.9in]{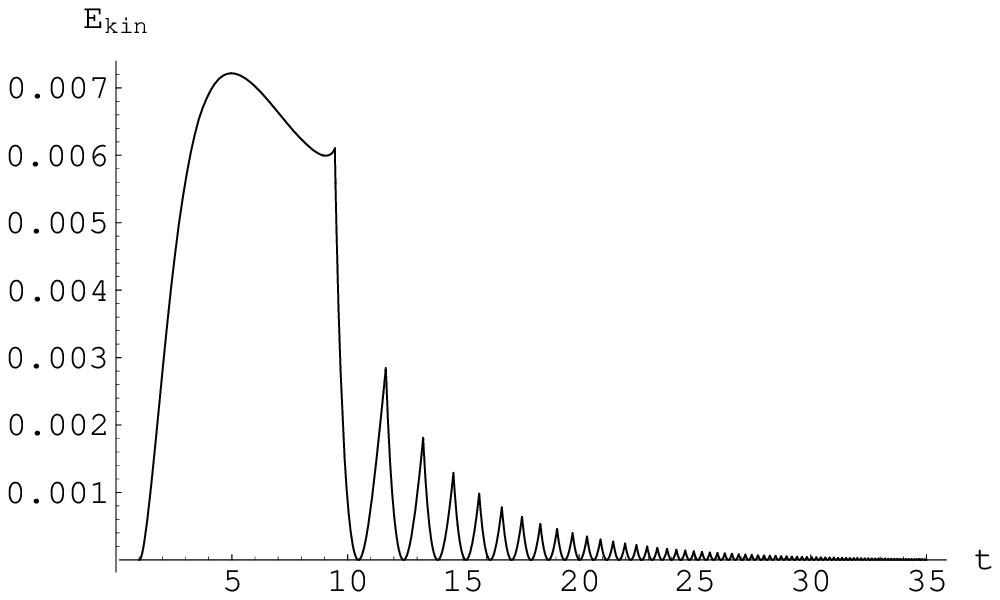}}
\caption {Dynamical behavior of scalar field $\theta (t)$ (left panel) and its
kinetic energy $E_{kin}(t)$ (right panel) in the vicinity of the branching point $\theta =2\pi$. \label{theta}}
\end{figure*}
We plot in Fig. \ref{H,a} the evolution of the logarithms of the
scale factor $a(t)$ (left panel) and the evolution of the Hubble
parameter $H(t)$ (right panel)
\begin{figure*}[htbp]
\centerline{\includegraphics[width=2.1in,height=1.9in]{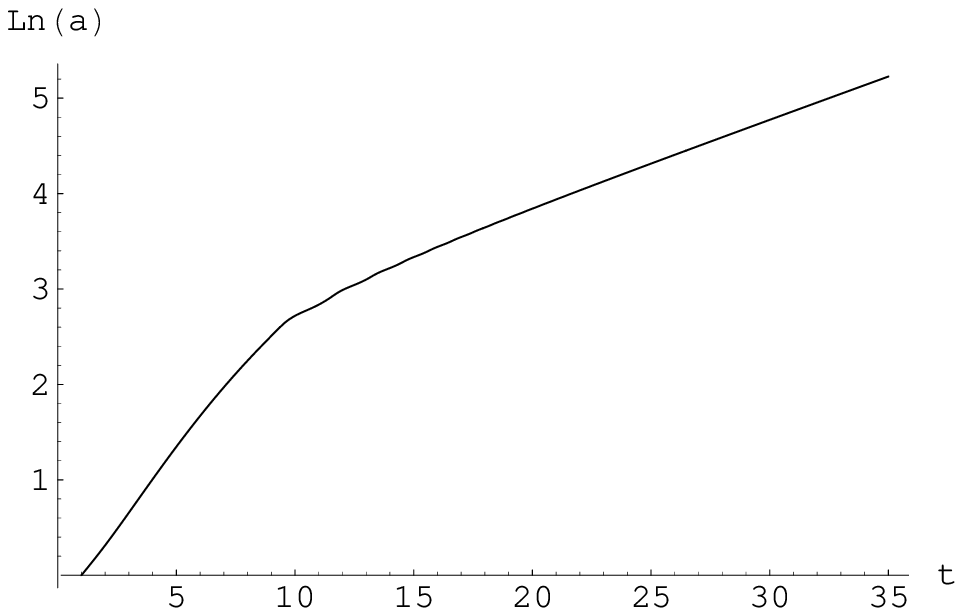}
\includegraphics[width=2.1in,height=1.9in]{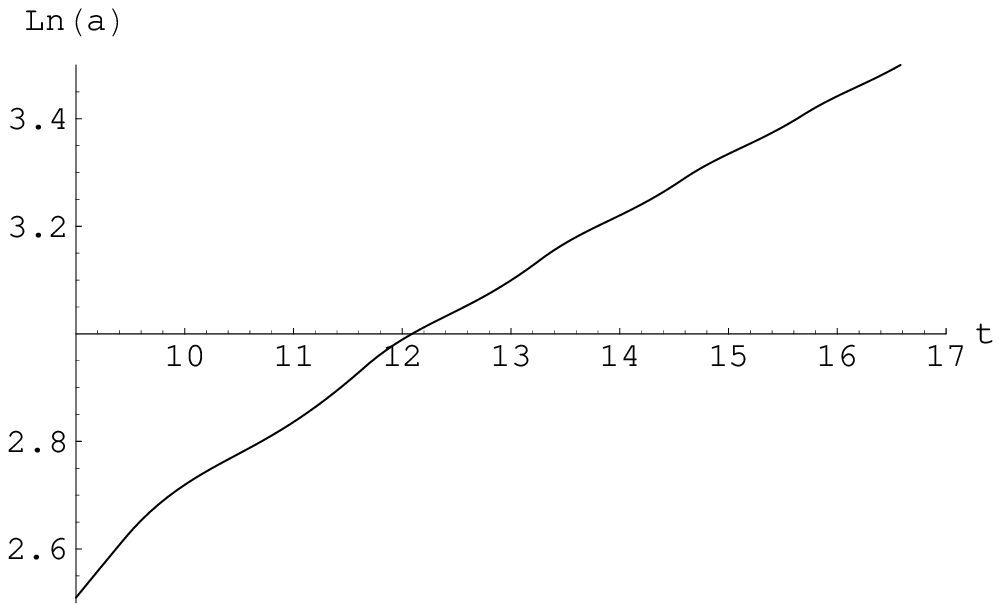}
\includegraphics[width=2.1in,height=1.9in]{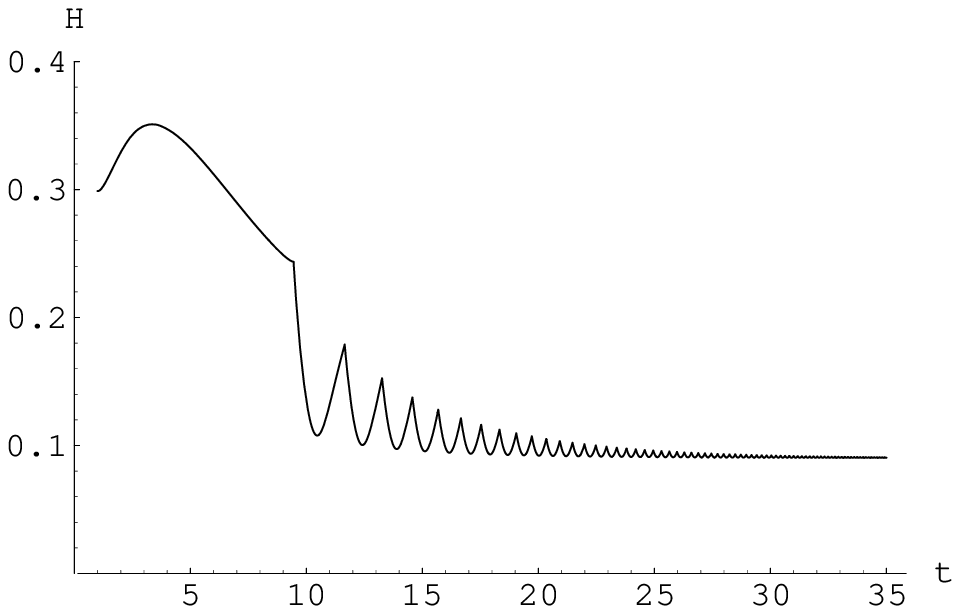}}
\caption {The time evolution of the logarithms of the scale factor $a(t)$ (left panel) and the Hubble
parameter $H(t)$ (right panel) for the trajectory drawn in Fig. \rf{theta}. A slightly visible oscillations of $\ln (a)$  (caused by bounces) can be seen  by magnification of this picture (middle panel).
\label{H,a}}
\end{figure*}
and in Fig. \ref{q} the evolution
of the parameter of acceleration $q(t)$ (left panel) and the equation of state parameter $\omega (t)= [2q(t)+1]/3 $ (right panel).

Fig. \rf{theta} demonstrates that scalar field $\theta$
bounces  an infinite number of times with decreasing amplitude
in the vicinity of the branching point
$\theta =2\pi$. $\theta$ cannot cross this point.  From Figs. \rf{H,a} and \rf{q} we see that the Universe asymptotically approaches the de Sitter
stage: $H\rightarrow \const, \quad q \rightarrow +1, \quad$ and $\omega \rightarrow -1$. Such accelerating behavior we call {\em bouncing inflation}.

Concerning the dynamical behavior in the vicinity of the branching point $\theta =\pi$, our analysis (similar performed above) shows that the scalaron field
$\theta $ cannot cross this local maximum regardless of the magnitude of initial velocity in the direction of $\theta=\pi$. It bounces back from this point.

\subsubsection{monotonic points
$\theta =0, 3\pi$}

Now, we want to investigate the dynamical behavior of the model in the vicinity of the monotonic points $\theta=0,3\pi$ which are the points
of inflection of the potential in Fig. \rf{pot}. Figs. \rf{pot} and \rf{force-mass} show that for both of these points
the model has the similar dynamical behavior. Therefore, for definiteness, we consider the point $\theta=3\pi$.
To investigate numerically the crossing of the monotonic point $3\pi$, it is necessary to take very small value of a step $\Delta t$. It can be achieved
if we choose very large value of the maximum number of steps. Thus, for the given value of the maximum number of steps, the closer to $3\pi$ the initial value $\theta_{initial}$ is taken the smaller step $\Delta t$ we obtain. For our calculation we choose $\theta_{initial}=9.513$.
Fig. \rf{theta 3P} demonstrates that scalar field $\theta $ slowly crosses the monotonic point $3\pi$ with nearly zero kinetic energy\footnote{The derivative $d\theta/dt$ goes to $-\infty$ when $\theta \to 3\pi$ (with different speed on  different sides of $3\pi$) but $d\phi/d\theta =0$ at $3\pi$ and kinetic energy is finite (see Eq. \rf{5.26}).\label{theta 3Pi}}. Then, just after the crossing,
the kinetic energy has its maximum value and starts to decrees gradually when $\theta$ moves to the direction $2\pi$.
\begin{figure*}[htbp]
\centerline{\includegraphics[width=2.1in,height=2.0in]{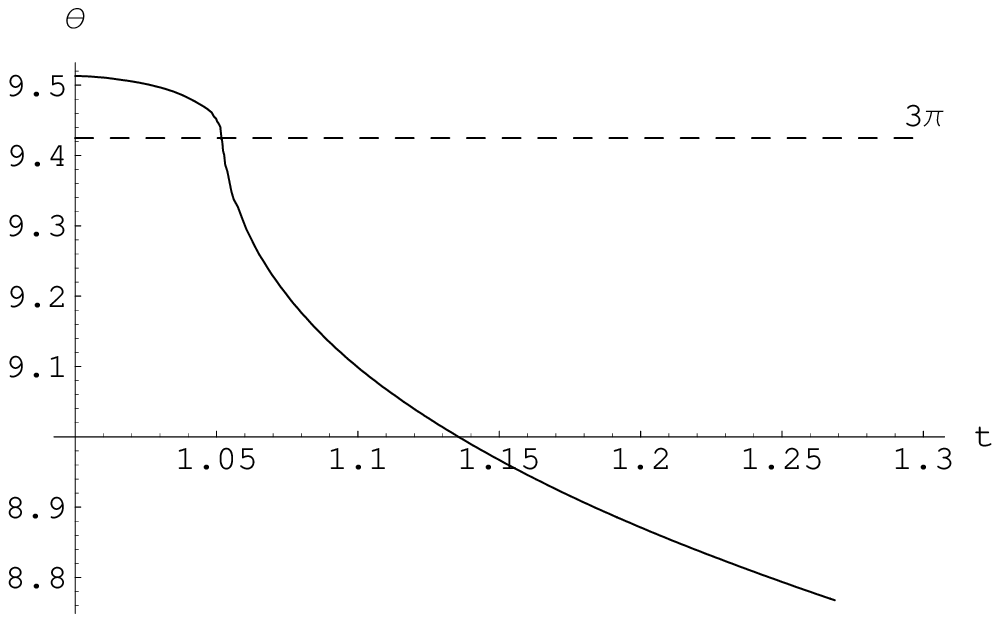}
\includegraphics[width=2.1in,height=2.0in]{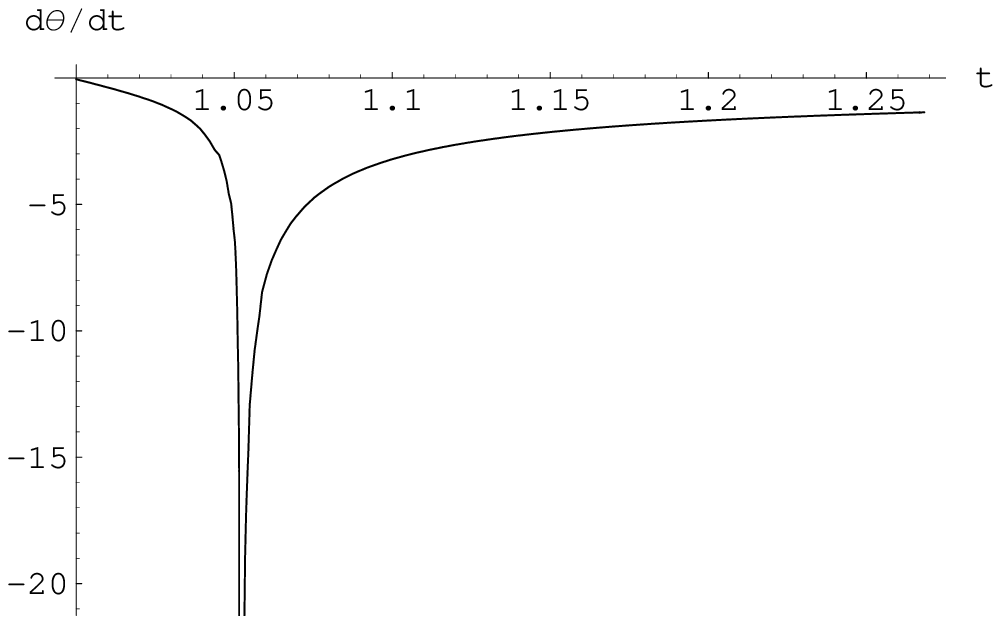}
\includegraphics[width=2.1in,height=2.0in]{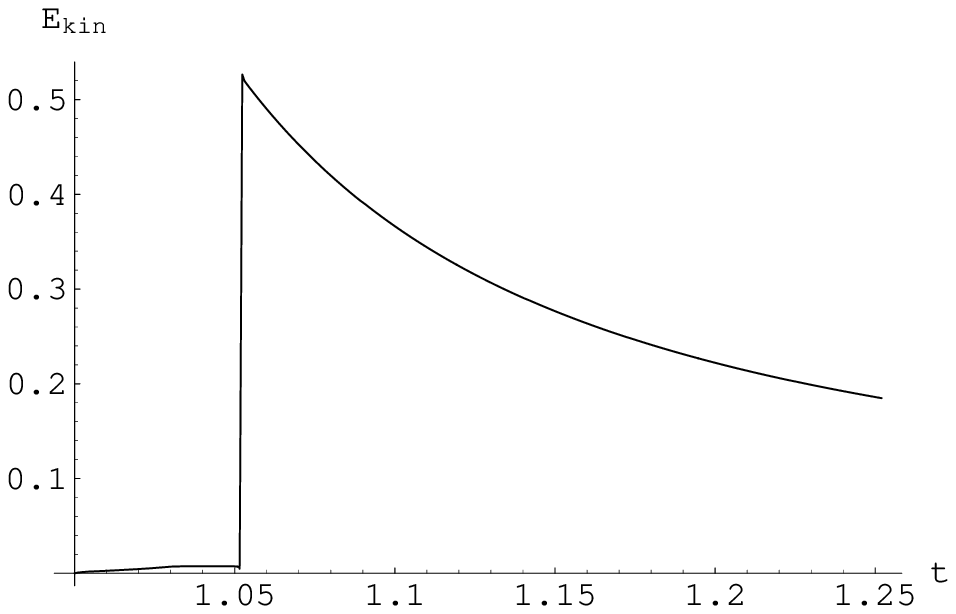}}
\caption {Dynamical behavior of scalar field $\theta (t)$ (left panel) and its time derivative $d\theta /dt$ (middle panel)
and kinetic energy $E_{kin}(t)$ (right panel) for the case of crossing of the inflection point $\theta =3\pi$.
\label{theta 3P}}
\end{figure*}
\begin{figure*}[htbp]
\centerline{\includegraphics[width=3.0in,height=2.0in]{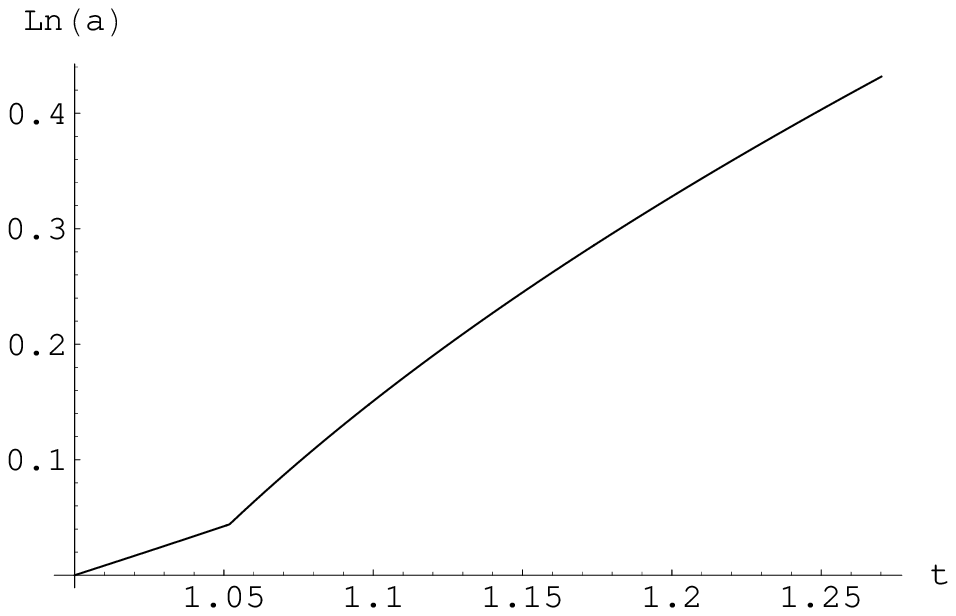}
\includegraphics[width=3.0in,height=2.0in]{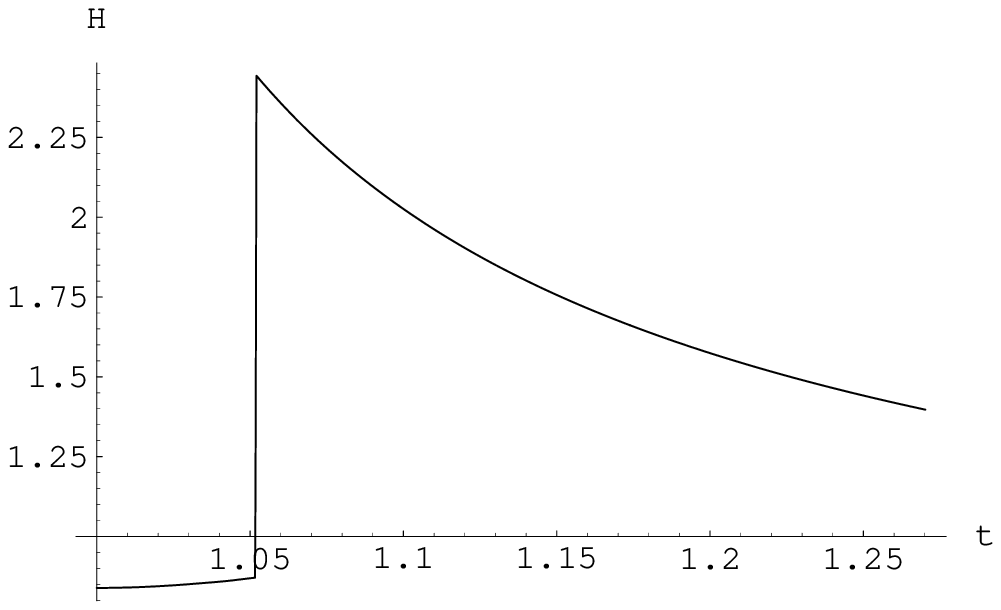}}
\caption {The time evolution of the logarithms of the scale factor $a(t)$ (left panel) and the Hubble
parameter $H(t)$ (right panel) for the trajectory drawn in Fig. \rf{theta 3P}.
\label{H,a3P}}
\end{figure*}
\begin{figure*}[htbp]
\centerline{\includegraphics[width=3.0in,height=2.0in]{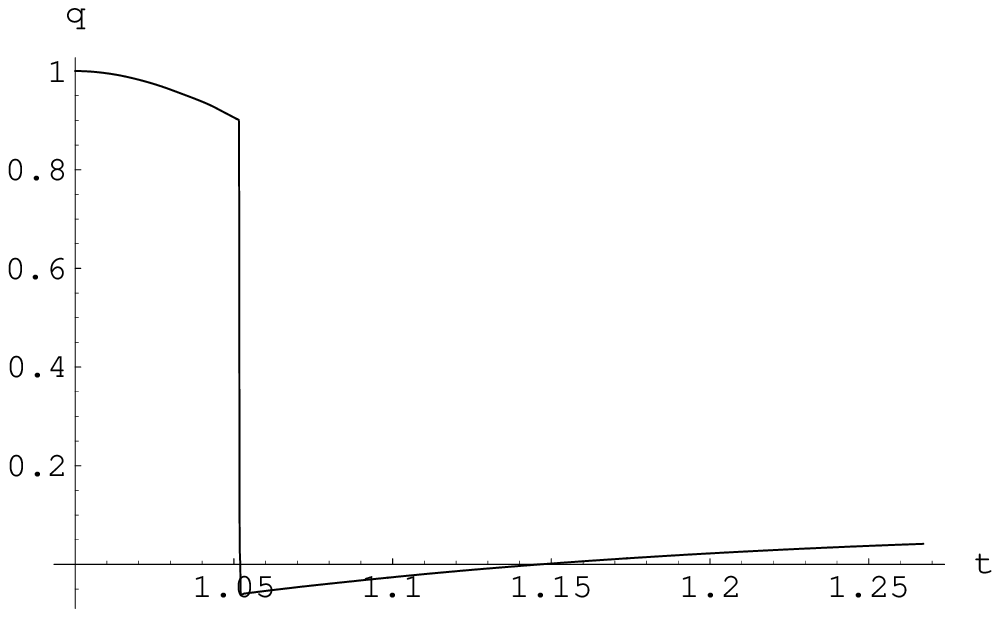}
\includegraphics[width=3.0in,height=2.0in]{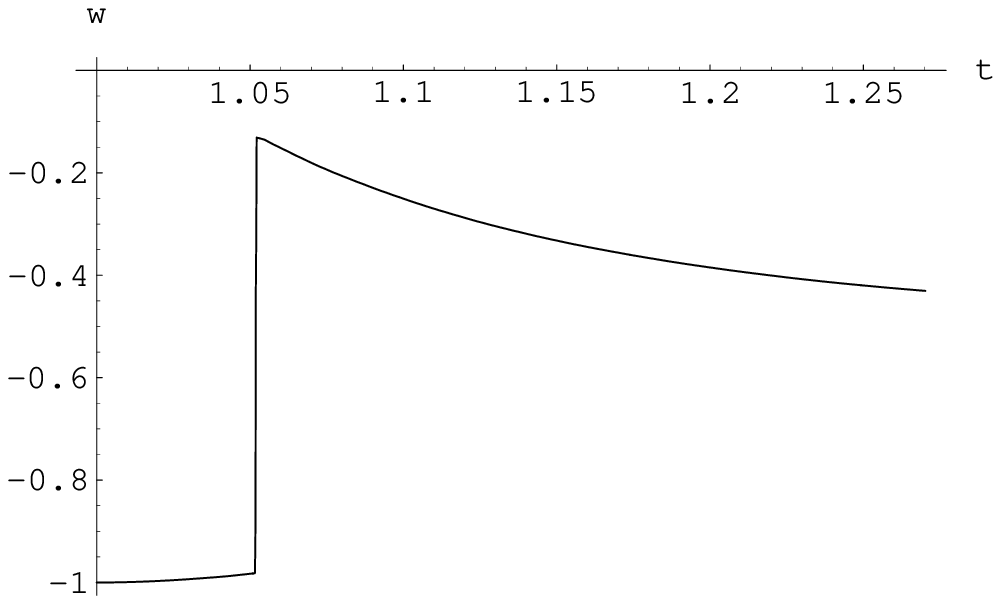}}
\caption {The parameter of acceleration $q(t)$ (left panel) and the equation of state parameter $\omega (t)$ (right panel) for the scale factor in Fig. \rf{H,a3P}.
\label{q3P}}
\end{figure*}
\begin{figure*}[!]
\centerline{\includegraphics[width=3.0in,height=2.0in]{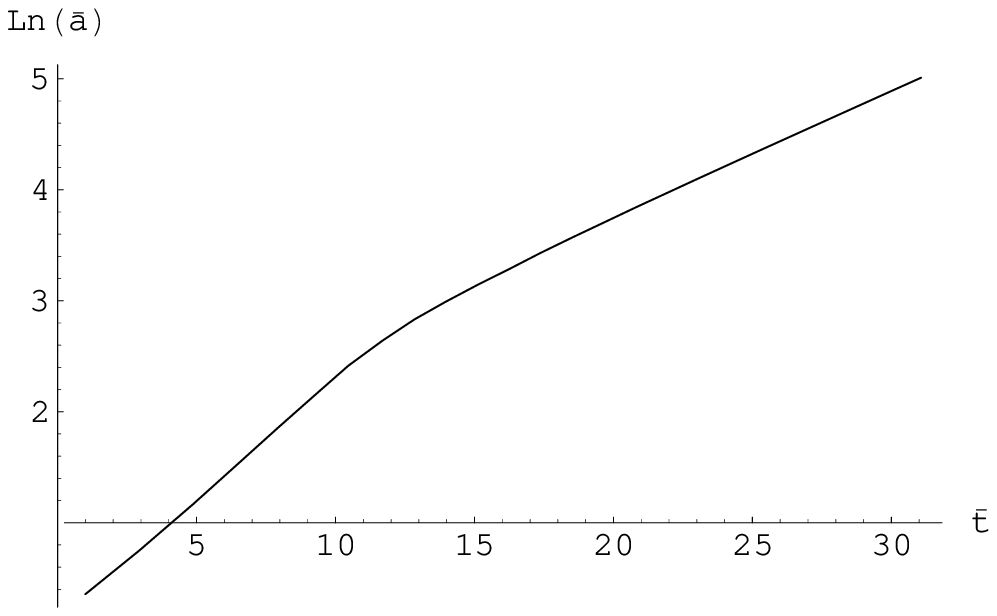}
\includegraphics[width=3.0in,height=2.0in]{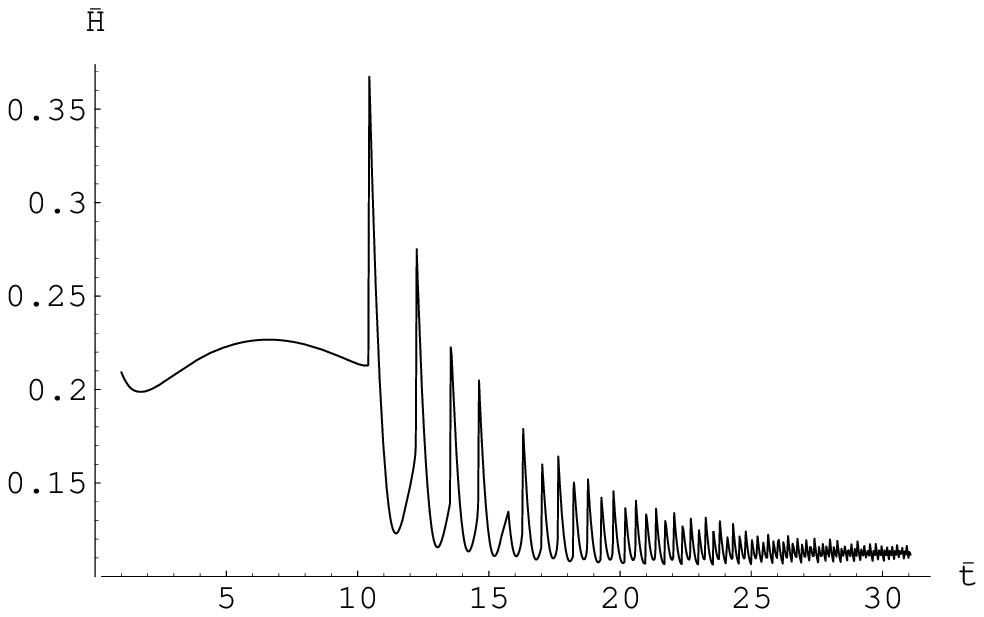}}
\caption {The time evolution of the logarithms of the scale factor $\bar a(\bar t)$ (left panel) and the Hubble
parameter $\bar H(\bar t)$ (right panel) for the trajectory drawn in Fig. \rf{theta} in the Brans-Dicke frame.
\label{BD}}
\end{figure*}
Figs. \rf{H,a3P} and \rf{q3P} demonstrate the behavior of the Universe before and after crossing $3\pi$. We do not show here the vicinity of the branching point $2\pi$ because when $\theta$ approaches $2\pi$ the Universe  has the bouncing inflation described above. Hence, there are
3 phases sequentially: the short de Sitter-like stage during slow rolling in the vicinity of the
inflection point before crossing, then decelerating expansion just after the crossing with gradual transition to the accelerating stage again when $\theta$ approaches
the branching point $2\pi$.
Clearly that for another monotonic point
 $\theta=0$ we get the similar crossing behavior (without the bouncing stage when $\theta \to -\infty$). Therefore, the monotonic
points $\theta=0$ and $\theta=3\pi$ are penetrable for the scalaron field $\theta$.
\section{Conclusions}

We have investigated the dynamical behavior of the scalaron field $\phi$ and the Universe in nonlinear model
with curvature-squared and curvature-quartic correction terms: $f(\bar{R})=\bar{R}+\alpha\bar{R}^{2}+\gamma\bar{R}^{4}-2\Lambda_{D}$. We have chosen parameters $\alpha$ and $\gamma$ in such a way that the scalaron potential $U(\phi )$ is a multi-valued function consisting of a number of branches. These branches are fitted with each other either in the branching points (points $P_{2,3}$ in Fig. \rf{U(X)}) or in the monotonic points (points $P_{1,4}$ in Fig. \rf{U(X)}). We have reparameterized the potential $U$ in such a way that it becomes the one-valued function of a new field variable $\theta =\theta(\phi)$ (see Fig. \rf{pot}). This has enabled us to consider the dynamical behavior of the system in the vicinity of the branching and monotonic points in ($D=4$)-dimensional space-time. Our investigations show that the monotonic points are penetrable for scalaron field (see Figs. \rf{theta 3P}-\rf{q3P}) while in the vicinity of the branching points scalaron has the bouncing behavior and cannot cross these points. Moreover, there are branching points
where scalaron bounces  an infinite number of times with decreasing amplitude and the Universe asymptotically approaches the de Sitter stage (see Figs. \rf{theta}-\rf{q}). Such accelerating behavior we call bouncing inflation. It should be noted that for this type of inflation there is no need for original potential $U(\phi)$ to have a minimum or to check the slow-roll conditions. A necessary condition is the existence of the branching points. This is a new type of inflation. We show that this inflation takes place both in the Einstein and Brans-Dicke frames. We have found this type of inflation for the model with the curvature-squared and curvature-quartic correction terms which play an important role during the early stages of the Universe evolution. However, the branching points take also place in models with $\bar R^{-1}$-type correction terms \cite{Frolov}. These terms play an important role at late times of the evolution of the Universe. Therefore, bouncing inflation may be responsible for the late-time accelerating expansion of the Universe.

To conclude our paper, we want to make a few comments. First, there is no need for fine tuning of the initial conditions to get the bouncing inflation. In Figs. \rf{theta}-\rf{q}, we have chosen for definiteness the initial conditions $\theta =3.5$ and $E_{kin}=0$. However, our calculations show that these figures do not qualitatively change if we take arbitrary $\theta \in (\pi,2\pi)$ and non-zero $E_{kin}$. Second, Figs. \rf{force-mass} indicates that the minimum at $\theta =2\pi$ is stable with respect to tunneling through the barrier at this point. The situation is similar to the quantum mechanical problem with infinitely high barrier. We have already stressed that the form of the potential Fig. \rf{pot} is not sufficient to predict the dynamical behavior of $\theta$. This field has very non-trivial behavior because of non-canonical kinetic term and singular (at the matching points) non-flat moduli space metric $G(\theta )$. Therefore, it is impossible to "jump" quantum mechanically from one branch to another. We cannot apply to our dynamical system the standard tunneling approach (e.g., in \cite{Coleman}). This problem needs a separate investigation. Third, it is worth noting that the Universe with a bounce preceding the inflationary period was considered in \cite{Yifu} where it was shown that due to a bounce the spectrum of primordial perturbations has the characteristic features. It indicates that the similar effect can take place in our model. This is an interesting problem for future research.

\mbox{} \\ {\bf Acknowledgments}\\ We want to thank Uwe G\"unther for useful comments.
We also would like to thank Yi-Fu Cai for drawing our attention to their paper \cite{Yifu}. 
This work was supported in part by the "Cosmomicrophysics" programme of the Physics and Astronomy
Division of the National Academy of Sciences of Ukraine.
\appendix
\section{\label{sec:A}Bouncing inflation in the Brans-Dicke frame}
\renewcommand{\theequation}{A.\arabic{equation}}
\setcounter{equation}{0}

According to Eq. \rf{1.9}, the four-dimensional FRW metrics in the Einstein frame \rf{5.1} and in the Brans-Dicke frame
are related as follows:
\be{a1}
-dt\otimes dt + a^2(t)d\vec{x}\otimes d\vec{x} =
f'\left[-d\bar t\otimes d\bar t + \bar a^2(\bar t)d\vec{x}\otimes d\vec{x}\right],
\ee
where $f'=X+1>0$ and $X$ is parameterized by Eq. \rf{5.3}.
Therefore, for the synchronous times and scale factors in both frames we obtain, correspondingly:
\ba{}
d \bar t &=& d t /\sqrt{f'(t)} \label{a2}\, ,\\
\bar a ({\bar t}) &=& a \left(t(\bar t)\right)/\sqrt{f'\left(t(\bar t)\right)}\label{a3} ,
\ea
which lead to the following equations:
\be{a4}
\bar t = \int^t_1\frac{d t}{\sqrt{X(t)+1}} +1\, ,
\ee
where we choose the constant of integration in such a way that $\bar t (t=1)=1$, and
\ba{a6}
\bar H (\bar t) &=& \left.\frac{d\bar a}{d\bar t} \frac{1}{\bar a}= \sqrt{X\left(t(\bar t)\right) +1}\right[H\left(t(\bar t)\right) \nn\\&-&\left.
\frac{1}{2(X\left(t(\bar t)\right)+1)}\frac{dX}{dt}\left(t(\bar t)\right)\right]\, .
\ea
From the latter equation we get the relation between the Hubble parameters in both frames.
We plot in Fig. \rf{BD} the logarithms of the scale factor $\bar a(\bar t)$ and the Hubble parameter $\bar H (\bar t)$
for the trajectory drawn in Fig. \rf{theta}. These pictures clearly demonstrate that in the Brans-Dicke frame the Universe has also asymptotical de Sitter
stage when the scalaron field approaches the branching point $\theta = 2\pi$.
It is not difficult to verify that because $X(t\to +\infty) \to X_{max}$ and $dX/dt(t\to +\infty) \to 0$,
we obtain the following relation for the asymptotic values of the Hubble parameters in both frames:
\be{a7}
\bar H = H \sqrt {X_{max}+1}\, .
\ee

\end{document}